\documentclass[12pt]{article}
  
\usepackage{amsmath,amssymb}

\usepackage{epsfig}  
\usepackage{graphicx}               % Standard graphics package  
\usepackage{url}

\newcommand{\nc}{\newcommand}  

\nc{\beq}{\begin{equation}}  
\nc{\eeq}{\end{equation}}  
\nc{\beqa}{\begin{eqnarray}}  
\nc{\eeqa}{\end{eqnarray}}  
\nc{\bea}{\begin{eqnarray}}  
\nc{\eea}{\end{eqnarray}}  
\nc{\ra}{\rightarrow}  
\nc{\lsim}{\begin{array}{c}\,\sim\vspace{-21pt}\\< \end{array}}  
\nc{\gsim}{\begin{array}{c}\sim\vspace{-21pt}\\> \end{array}}  
\nc{\Tr}{{\rm Tr}}
\nc{\slsh}{\slash\hspace*{-0.22cm}}

\makeatletter
\def\url@leostyle{%
  \@ifundefined{selectfont}{\def\UrlFont{\sf}}{\def\UrlFont{\small\ttfamily}}}
\makeatother
\title{  
\vspace*{-2.3cm}  
\begin{flushright}  
\normalsize{  
FERMILAB-PUB-09-527-T\\ 
  }  
\end{flushright}  
\vspace{1.5cm}  
\Large  
\textbf{Topological Interactions in Warped Extra Dimensions
}\vspace*{1.0cm}   
}
\author{\large
\textbf{Yang Bai,$^a$} 
\textbf{Gustavo Burdman,$^{a,b}$}  
\textbf{Christopher T. Hill$^{a}$}
\\ \\[0.5cm]  
$^a$\normalsize\emph{Fermi National Accelerator Laboratory,  
P.O. Box 500, Batavia, IL 60510, USA} \\  
$^b$\normalsize\emph{Instituto de F\'{i}sica, Universidade de S\~{a}o Paulo, Brazil}
}
\date{}

\begin{document}  
\setcounter{page}{0}  
\maketitle  

\vspace*{0.5cm}  
\begin{abstract} 
\vspace*{0.5cm}
Topological interactions will be generated in theories with compact extra dimensions where fermionic chiral zero modes have different localizations. This is the case in many warped extra dimension models where the right-handed top quark is typically localized away from the left-handed one. Using deconstruction techniques, we study the topological interactions in these models.  These  interactions appear as trilinear and quadrilinear gauge boson couplings in low energy effective theories with three or more sites, as well as in the continuum limit. We derive the form of these interactions for various cases, including examples of Abelian, non-Abelian and product gauge groups of phenomenological interest. The topological interactions provide a window into the more fundamental aspects of these theories and could result in  unique signatures at the Large Hadron Collider, some of which we explore. 
\end{abstract}  

\thispagestyle{empty}  
\newpage  
  
\setcounter{page}{1}

\baselineskip18pt   

\section{Introduction} 
The origin of electroweak symmetry breaking (EWSB) is one of the most 
important questions in particle physics and will likely lead to the 
discovery of new organizing principles beyond the standard model. 
As we enter the era of the Large Hadron Collider (LHC) with the promise
of new discoveries of new states in nature,  
it remains nonetheless unclear how much of the deeper organizing context 
for EWSB can be understood at the TeV scale.
 
For instance, consider solutions of the gauge
hierarchy problem involving theories with a compact extra dimension in AdS space~\cite{rs1}. 
These are thought to be a dual description of a large-$N$, $D=4$, strongly 
coupled sector characterized by conformal dynamics~\cite{holo,adscft}.  
The low energy spectrum, however, is typically populated by an assortment
of new vector resonances with various standard model quantum numbers, 
and possibly new heavy fermions.   We might then
seek probes that could reveal the deeper UV completion
structure.  These can arise from anomalous, or
``topological" processes, associated with the gauge dynamics of
chiral fermions, much like the low energy process, $\pi^0\rightarrow 2\gamma$, 
counts the quark colors in QCD.  

Chiral fermions are required as part  
of the low energy spectrum of any model, often
arising by chiral localizations in extra dimensional models,
whereby left-handed fermions
occur at one place in the bulk, whilst their right-handed anomaly-canceling partners 
occur elsewhere.  This has immediate implications for the anomaly structure 
of such theories, or more properly, the Chern-Simons (CS) term structure. 
The CS term propagates the anomaly from one chiral fermion to another and maintains the
anomaly cancellation across the extra dimension.
Although it is well understood that anomalies in 
orbifold theories are brane-localized and canceled 
by a suitable bulk CS term, the associated observable consequences  
of CS terms have not been fully elaborated in the literature. 
This is important since the associated CS interactions, involving gauge KK-modes,
point to fundamental aspects of the underlying theory in analogy to $\pi^0\rightarrow 2\gamma$
in QCD.   

Anomalies and CS terms in extra dimensional models have been previously considered in the 
literature.  These descend from gauge boson solitons that are topological objects, such as the instantonic vortex, arising
in $D=5$ and whose conserved currents are generated by the CS term \cite{D5vort}.  
In Ref.~\cite{Skiba:2002nx}, deconstruction was used to show how 
anomalies are canceled in theory space by 
the Wess-Zumino-Witten (WZW) terms present on each link, 
as well as illustrating the appearance of the CS term in the 
continuum limit of a flat compact extra dimension (see also ~\cite{Hill:2002me}). To cancel anomalies in orbifold theories with 
delocalized chiral zero modes, CS terms must necessarily occur. 
CS terms, in turn, produce physical consequences: in $D=3$ QED the CS
term yields a mass for the photon and destroys Dirac magnetic monopoles (see \cite{Hill:2009hx} and references therein).
Likewise, in $D=5$, CS terms lead to observable physical effects, first pointed out in Ref.~\cite{Hill:2006ei}, where, as a general consequence, 
they violate KK-parities. This is analogous to the violation
of $\pi \rightarrow -\pi$  spurious pion parity  
in a chiral lagrangian of mesons by the WZW term
in QCD.  These physical effects can most easily be understood by considering  three
very massive $D=5$ bulk KK-mode wave-packets each vanishing on the branes 
where the fermions are localized. The bulk  CS term
operator will generally have a non-vanishing overlap integral for such 
wave-packets,
provided overall KK-mode parity is odd.
The pure CS term in the bulk controls these interactions. 
For lower KK modes, whose wave-functions touch the fermionic branes,
the loop diagrams of the localized fermions become relevant, leading to the
counterterm structure that enforces, e.g., vector-like current conservation \cite{Hill:2006ei}
(or, alternatively, chiral current conservation, with the appropriate counterterm \cite{HHH}).

This observation has been applied in Refs.\cite{Hill:2007zv} to Little Higgs theories,
which can be viewed as deconstructed extra dimensional theories. Aside from
identifying certain special processes in electron or muon collider experiments
that probe CS terms, a key result is that it  is generally difficult
to maintain a stable dark matter KK-mode candidate in the presence of CS terms.
This is an
effect that will recur in the present paper.  
Some aspects of anomalies in warped extra dimension models were studied 
in Ref.~\cite{gripaios}.

In this paper we will consider the remnant topological interactions 
at low energy resulting from bulk Chern-Simons terms in theories 
with warped extra dimensions. In order to clarify the origin of these 
new interactions amongst KK gauge bosons we first 
deconstruct~\cite{decgen} these theories (this was previously done in Refs. \cite{decfermions,decgauge}).  Pure gauge boson
containing CS-term interactions are seen to be absent in two-site deconstructions\footnote{
This is strictly true for vector-like gauge zero-modes.
A two site model is analogous to the chiral constituent quark
model $U(N)_L\times U(N)_R$ with quarks $q_L$ and gauge bosons $A_L$
($q_R$ and $A_R$) on the L-site (R-site), and a constituent quark mass involving
pions. This can be viewed as descending from a
vector-like $SU(N)$ $D=5$ Yang-Mills bulk theory with chiral localizations generated by domain walls.
A CS term is present in $D=5$ and becomes
the WZW term in $D=4$ that compensates the quarks consistent anomalies on the 
walls. The WZW term contains trilinear and quadrilinear "pCS" terms 
\cite{HHH} at this stage, but when the quarks are integrated out, a Bardeen counterterm is
generated, conserving the vector currents, enforcing the Landau-Yang theorem and annihilating the pCS terms.  
In  a three site model even with quarks integrated out and conserved zero-mode vector
currents, the pCS terms involving higher KK-modes remain. Note that one can modify the counterterm when
the $L$, or $R$ currents are conserved, as in Standard Model gauging. This is then not a vector-like scheme.}.
These first appear in deconstructions with three or more sites. 
This  has important consequences in the phenomenology of these 
interactions in the continuum limit, 
most notably the fact that -- as long as the zero-mode gauge symmetry remains unbroken -- these
interactions must involve the second KK mode
of the gauge boson. Amongst the various interactions the most accessible involves the 
gluon and its first and second KK modes. 
This  results in a new decay mode for the second KK mode: $ G^{(2)}\to G^{(1)} g$, which gives a 
non-negligible contribution to the $G^{(2)}$ width.  It also allows for the associated 
production process $pp\to G^{(2)} G^{(1)}$ to be observable at the LHC as long 
as the KK masses are not too heavy, as is the case, for instance, in warped Higgsless models.

We  will show that the requirement that a second KK mode be present 
can be circumvented when one of the gauge bosons in the interaction
is associated with a broken gauge symmetry. This leads to many new interactions 
involving the $Z$ boson with KK gauge bosons and 
other zero modes. We will study the phenomenology of the most promising 
interactions, including the one involving a gluon and 
its two first KK  modes, as well as one with a gluon, its first KK mode and the $Z$.  

Finally, we consider a proposed warped extra dimension scenario with KK parity \cite{Agashe:2007jb} 
and show that the topological interactions do not  
break this symmetry, and still allow for a stable dark matter candidate in the lightest KK-odd particle. 
 
In the next section we consider in detail the
deconstruction \cite{decgen} of a warped $D=5$ theory including fermions and gauge bosons. 
A complete  treatment of fermions in warped extra dimension theories is not present in the literature
and is central for our derivations. 
In deconstruction, the CS term becomes a sum over interlinking WZW terms.
In unitary gauge, where all KK-modes eat their corresponding Nambu-Goldstone Bosons (the link field phases) the
sum over WZW terms immediately reduces to the discretized version of the CS term. None of this makes any sense, however, without
utilizing Wilson fermions in deconstructed theories. 
We make use of the Wilson fermion action for warped theories, first introduced in Refs.~\cite{Hill:2002me,decfermions}, 
and further developed here. 
It is crucial for our derivation of the WZW terms which give rise to the bulk Chern-Simons terms in the continuum limit. 
This is done in Section~\ref{wzwtocs}, in the limit of extreme zero-mode fermion localization,
Ref.~\cite{Agashe:2007jb}, where we also explicitly show how anomaly cancellation 
works. 
In Section~\ref{ioutfermions} we show that the existence and detailed form of the 
induced topological interactions depend on the localization of zero-mode fermions.  
In Section~\ref{topproc} we derive the remnant processes of interest for phenomenological 
applications: 
interactions among three vector states 
involving Kaluza-Klein modes of the gauge bosons. 
We show how these interactions arise from the Chern-Simons terms paying particular attention 
to gauge invariance. Finally, in  Section~\ref{pheno} 
we study some of the phenomenological consequences of these interactions,  
such as collider signals at the LHC, as well as the induced breaking of KK parity in
models with a $Z_2$-symmetric warp factor. 
We conclude in Section~\ref{conclusions}.

%--------------------------------------------------------------
\section{Deconstruction of a  Warped Extra Dimension} 
%--------------------------------------------------------------
\label{decon}
In order to clarify the presence of remnant topological interactions in theories 
with warped extra dimensions and chiral zero modes, we will deconstruct the extra dimension~\cite{decgen}. 
The deconstruction of warped extra dimensions has been studied for the gauge sector in Ref.~\cite{decgauge}, whereas fermions are also 
considered in Ref.~\cite{decfermions}. 
We first briefly review the warped extra dimension scenario in the continuum. 

We start with one extra dimension $y$ compactified on an orbifold $S_1/Z_2$, with $-L\leq y\leq L$, and  with the metric~\cite{rs1}
\beq
ds^2 = e^{-2ky}\eta_{\mu\nu}dx^\mu dx^\nu - dy^2
\equiv g_{MN}dx^Mdx^N
\eeq
where $\mu=0,1,2,3$, $k$ is the AdS$_5$ curvature and $\eta_{\mu\nu}={\rm diag}(+\,-\,-\,-)$ 
is the 4D Minkowski metric. In the following, we will use Greek letters for 4D indexes and Latin letters for 5D indexes. 

For fermions and gauge bosons propagating in the bulk of AdS$_5$, 
the 5D action is then given by~\cite{bulkwarped}
\beqa
S_5\,=\,\int d^4x  \int^L_0 dy \sqrt{g}
\left[ -\frac{1}{2\,g_5^2}{\rm Tr}[F^2_{MN}]\,+\, i\,\bar{\Psi}\,
\Gamma^M\nabla_M\,\Psi \,+\,M_\Psi\,\bar{\Psi}\Psi \,+\,\cdots \right]\,.
\label{actionall}
\eeqa
Not shown are the 5D Ricci scalar and the cosmological constant.
The fifth dimension $y$ is compactified with the IR (UV) 
branes located at the $y = L\,(0)$ of the fifth dimension. $F_{MN}$ 
is the field strength of the gauge group, which can be either Abelian 
or non-Abelian. The gamma matrices are defined as $\Gamma_M = e^A_M \gamma_A$, 
where $e^A_M$ is the vielbein, and $\gamma_A = 
(\gamma_\alpha,i \gamma_5)$ is defined in the tangent space. The curved space covariant 
derivative is $\nabla_M = D_M + \omega_M$, with the spin connection 
$\omega_M = (\frac{k}{2}\,\gamma_5\,\gamma_\mu e^{-ky}, 0)$. 
The fermion Dirac mass is $M_\Psi \equiv c\,k$, and is assumed to be the result of the vacuum expectation value of a 
scalar field odd under a $Z_2$ transformation defined by $y\to -y$.

Before deconstructing this model, we review the spectrum of KK modes and their 5D wave 
functions, both for gauge bosons and 
fermions. 
By choosing Neumann-Neumann boundary conditions for the gauge boson in the following way: 
$\partial_5 A_\mu(0) = \partial _5 A_\mu(L) = 0$ and $A_5(0) = A_5(L) =0$, 
the $A_\mu$ has a 4D zero mode with a flat profile in the fifth dimension. 
The equation of motion for the massive Kaluza-Klein (KK) modes is~\cite{bulkwarped}
\beqa
\partial_5^2 f_n - 2\,k\,\partial_5 f_n + m_n^2\,e^{2\,k\,y} f_n = 0\,,
\eeqa
where the KK expansion is given by
\beq
A_\mu(x, y) \equiv \frac{1}{\sqrt{L}}\,\sum^\infty_{n=0} f_n(y) A^n_\mu(x)\,, 
\qquad A_5(x, y) \equiv \frac{1}{\sqrt{L}}\,\sum^\infty_{n=1}\frac{\partial_5 
f_n(y)}{m_n} A_5^n(x)\,,
\eeq
with the normalization condition $g_5^{-2}\int^L_0 f^2_n dy= 1$. The solution of the 
gauge boson KK modes is 
\beqa
f_n(y)\,=\,\frac{e^{k\,y}}{N_n}\,\left[J_1\left(\frac{m_n}{k\,e^{-ky}}\right) 
+ b_1 Y_1\left(\frac{m_n}{k\,e^{-ky}}\right)     \right]\,,
\label{eq:gaugeCont}
\eeqa
where $b_1$ is a function of the KK mode mass $m_n$ and is determined by the boundary 
conditions, and $N_n$ is a normalization factor.

Fermions must transform under the $Z_2$ symmetry  
as $\Psi(-y) = \pm \gamma_5 \Psi(y)$ with $\gamma_5 = {\rm diag} (1, -1)$. 
In terms of Dirac spinors $\Psi = \Psi_R + \Psi_L$, the zero mode of $\Psi_R (\Psi_L)$ 
is even for $\Psi(-y) = +\gamma_5 \Psi(y)$ ($\Psi(-y) = - \gamma_5 \Psi(y)$). 
Therefore, the choice of the boundary condition makes the low energy effective 
4D theory chiral. The equation of motion for the fermion KK modes is given by
\beq
\partial^2_5 h_{L,R}^n - 2\,k\,\partial_5\,h_{L,R}^n\,+\,\left(\frac{3}{4} - 
c (c\pm1)\right)k^2\,h_{L,R}^n\,+\, m^2_n\,e^{2ky}\, 
h_{L,R}^n  = 0\,,
\label{eq:fermionCont}
\eeq
with  ``+" for the left-handed modes and ``--" for the right-handed modes. Here, 
\beq
\Psi_{L,R} \equiv  \frac{e^{\frac{3}{2}\,k\,y}}{\sqrt{L}}\sum^\infty_{n=1} 
h_{L,R}^n(y)\,\psi^n_{L,R}(x)\,,
\eeq
and the normalization condition $\frac{1}{L}\int^L_0 \left|h^n_{L,R}\right|^2 dy = 1$. The solutions for the fermion KK modes are then 
\beqa
h_{L,R}^n(y)\,=\,\frac{e^{k\,y}}{N_n}\,\left[J_{|c\pm \frac{1}{2}|} 
\left(\frac{m_n}{k\,e^{-ky}}\right) + b_{|c\pm \frac{1}{2}|}Y_{|c\pm \frac{1}{2}|} \left(\frac{m_n}{k\,e^{-ky}}\right)     \right]\,.
\label{fermisol}
\eeqa
with $N_n$ normalization factors. 
The fermion zero modes have an exponential profile given by 
\beq
h_{L,R}^0(y)\,=\,\frac{1}{N_0}\,e^{(\frac{1}{2} \mp c)\,k\,y}\,.
\eeq
Therefore, a left-handed zero mode is UV (IR)-localized for $c_L >1/2$ $(c_L < 1/2)$.
On the other hand,  a right-handed zero mode is UV (IR)-localized for $c_R < -1/2$ 
$(c_R > -1/2)$. 
In a wide class of warped extra-dimension models, both left-handed and 
right-handed zero-modes of SM fermions are 
mostly UV-localized for all fermions except for the third generation quarks.
Typically in these models, in order to obtain a large enough top quark mass, 
$t_R$ is localized close to the IR brane, with the 
third generation quark doublet $(t_L ~b_L)^T$ somewhere in between the IR and 
UV branes. We will show later that it is precisely due to the different fifth-dimension 
profiles for the top quark chiral zero modes, that 
there exist physical topological interactions among gauge bosons.    

\subsection{Deconstruction of the 5D Gauge Theory and the Dictionary}
\begin{figure}[ht!]
\vspace{0.3cm}
\centerline{ \hspace*{0.0cm}
\includegraphics[width=0.8\textwidth]{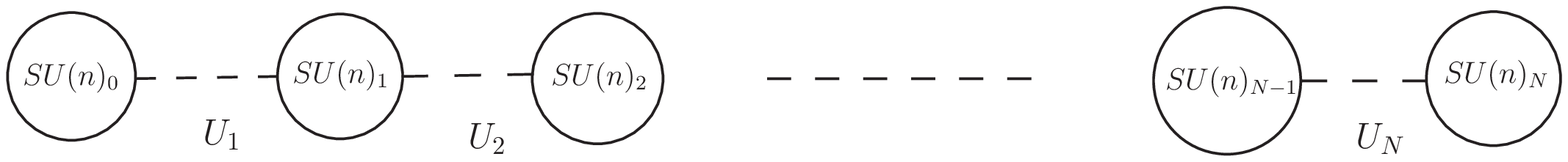}
}
\caption{\em Deconstruction of a gauge theory in a warped extra dimension. 
The circles represent $SU(n)$ gauge groups. The zeroth site and the $N$-th site are identified as the UV and IR brane in the continuous theory, respectively. 
The link scalar fields $U_j$, are $(n, \bar{n})$ under $SU(n)_{j-1}\times SU(n)_j$.  
\label{gaugemoose}
}
\end{figure}

In order to establish a dictionary between the continuum theory and the 
4D deconstructed one, we start with 
the purely bosonic 4D moose model with $N+1$ sites depicted in 
Figure~\ref{gaugemoose}. This results in the action
\beq
S_4^G = \,\int d^4 x\left\{-\frac{1}{2\,g^2}\,\sum_{j=0}^N \,\Tr \left[F^j_{\mu\nu}F^{j\mu\nu}\right]
+ \, \sum^N_{j=1} \Tr\,\left|D_\mu U_j
\right|^2
\right\} \,,
\eeq
with the covariant derivative given by $D_\mu U_j=\partial_\mu U_j + i\,A^{j-1}_{\mu}\,\,U_j  - i\,U_j\,A^{j}_{\mu}$ with $A^{j}_\mu \equiv A^{j}_{a, \mu}\,t^{a}$, where the link fields $U_j$ transform as $(n,\bar n)$ under $SU(n)_{j-1}\times SU(n)_j$ and $t^a$ is the generator of $SU(n)$ normalized as ${\rm tr}[t^at^b]=\frac{1}{2}\delta^{ab}$. We assume that the vacuum expectation values (VEVs) of $U_j$ break $SU(n)_{j-1}\times SU(n)_j$ to the diagonal group by minimizing some potentials. In the non-linear parametrization, we have $U_j = \frac{v_j}{\sqrt{2}}\,e^{i\,G_j/v_j}\,{\mathbb I}_n$, where the 
$G_j$ are the Nambu--Goldstone bosons of the breaking of  $SU(n)_{j-1}\times SU(n)_j$, 
and $v_j$ are the corresponding VEVs. Then, in the unitary gauge, 
we can write
\beq
S_4^G = \,\int d^4 x\left\{-\frac{1}{2\,g^2}\,\sum_{j=0}^N \,\Tr \left[F^j_{\mu\nu}F^{j\mu\nu}\right]
+\frac{1}{2}\,\sum_{j=1}^N \Tr\left[v_j(A_\mu^{j-1}-A_\mu^j)\right]^2
\right\}  \,.
\label{fourdgauge}
\eeq
In order to match to the continuum warped extra-dimension theory, we choose the VEV in each site as 
\beq
v_j\equiv v\,q^j  \,,
\label{eq:gaugevevs}
\eeq
such that $\langle U_j \rangle = \frac{v}{\sqrt{2}}\,q^j$ with $0< q <1$. 
Hence, from the zeroth site to the $N$-th site, the VEVs of the link field are decreasing. We identify the zeroth site as the UV brane and the $N$-th site as the IR brane when we match this discretized 4D model to the continuum warped space.

To justify the choice of the VEVs in Eq.~(\ref{eq:gaugevevs}), we need to show that the spectra and the wave-functions of the gauge bosons agree with the results in the continuum limit. For convenience, we choose the unitary gauge, in which the gauge boson mass matrix in the basis $(A_0, A_1, \cdots, A_N)$ can be written in powers of $q$ as   
\beqa
M^2_g\,=\,g^2\,v^2\,\left(  
\renewcommand{\arraystretch}{1.6}
\begin{array}{ccccccc}
q^2   &  -q^2   & 0               &   0   & \cdots    & 0   &    0     \\
-q^2  &  q^2 + q^4  &   -q^4     &  0  & \cdots  &   0    &0      \\
0    & -q^4     &  q^4 + q^6    &  - q^6  & \cdots & 0 & 0   \\
\vdots & \vdots & \vdots  & \vdots &  \cdots  & \vdots& \vdots   \\
0    & 0 & 0 &0 &\cdots &    q^{2(N-1)} + q^{2N}    &  -q^{2N}  \\
0 & 0 & 0&0 &\cdots &      -q^{2N}     & q^{2N} 
\end{array}
\right)\,.
\eeqa

We define the orthonormal rotation matrix between the gauge basis $A^n$ and mass basis $A^{(n)}$ as $A^j_\mu = \sum^N_{n=0} f_{j, n}\,A^{(n)}_\mu$. Solving this eigensystem problem, we arrive at the following difference equations~\cite{deBlas:2006fz}
\beqa
\left(q\,+\,q^{-1}\,-\,q^{-1}(x_n\,q^{-j})^2\right)\,f_{j, n}\,-\,q\,f_{j+1, n}\,-\,q^{-1}\,f_{j-1, n} \,=\,0\,,
\label{eq:gaugeEOM1}
\eeqa
The corresponding Neumann-Neumann  ``boundary conditions" are: 
$f_{0, n} = f_{-1, n}$ and $f_{N, n} = f_{N+1, n}$, with $x_n = m_n /(g\,v)$. 
For the gauge boson zero mode, it is easy to show that $f_{j, 0} = 1/\sqrt{N+1}$ ,  i.e. 
the solution is a  flat profile. 
For the massive modes, we define the variable $t[j] = x_n \,q^{-j}$ and 
the function $F(t[j]) = q^j\,f_{j, n}$ to change Eq.~(\ref{eq:gaugeEOM1}) to 
\beqa
(q\,+\,q^{-1}\,-\,q^{-1} t^2)\,F(t)\,-\,F(t\,q^{-1})\,-\,F(t\,q) \,=\,0\,.
\label{eq:gaugeEOM2}
\eeqa
The above difference equation is a special case of the  Hahn-Exton equation~\cite{hahnexton}. 
Its solutions are the so-called  $q$-Bessel functions $J_\nu(t; q^2)$ 
for $\nu=1$ in the mathematical literature. The solution of the difference equation in (\ref{eq:gaugeEOM2}) is 
\beqa
f_{j, n}\,=\,R_n\,q^{-j}\,\left[ J_1(x_n\,q^{-j}; q^2)\,+\,b_1(x_n; q^2)\,Y_1(x_n\,q^{-j}; q^2)               \right]\,,
\label{fjn}
\eeqa
with $R_n$ determined from wave-function normalization. This corresponds to the $j$-site ``wave-function'' of the $n$-th KK gauge boson, and it allows
us to construct the mass eigenstates $A^{(n)}_\mu$. Imposing the boundary conditions around $j=0$ and $j=N$ mentioned above, we obtain 
$b_1(x_n; q^2)$ and the following equation~\cite{deBlas:2006fz}
\beq
J_0(x_n; q^2)\,Y_0(x_n\,q^{-(N+1)}; q^2) - Y_0(x_n; q^2)\,J_0(x_n\,q^{-(N+1)}; q^2) = 0  \,,
\label{masscondition}
\eeq
the solution of which gives the mass spectrum. This procedure is very similar to the one followed in the continuum.
In fact it can be shown that in the continuum limit, corresponding to 
$q\rightarrow 1_-$, the solutions (\ref{fjn}) to the discrete equation of motion
match to the solutions (\ref{eq:gaugeCont})  for the wave-functions of the 
KK gauge bosons in the continuum. It is also easy to show that the mass eigenvalues
match to the KK-mode masses of the continuum theory. We can see the equivalence of both theories by using the following dictionary
\beqa
\frac{1}{g^2} \qquad &\leftrightarrow &\qquad\frac{a}{g_5^2} \,\\
v_j \qquad &\leftrightarrow & \qquad\frac{e^{-k a j}}{a} \,
\label{eq:dictionary}
\eeqa
We can then rewrite (\ref{fourdgauge}) as 
\beq
S_5^G = \,\frac{a}{g_5^2}\,\int d^4 x\,\left\{ -\frac{1}{2}\sum_{j=0}^N\,\Tr \left[F^j_{\mu\nu}F^{j\mu\nu}\right] 
+ \frac{1}{2}\sum_{j=0}^{N-1}\,e^{-2kaj}\,\,\Tr\left(\frac{A_\mu^{j+1}-A_\mu^{j}}{a}\right)^2\right\}\,,
\label{fivedgauge}
\eeq
where $a$ is the constant lattice spacing, and $g_5$ is the 5D gauge coupling. 
With these replacements and taking the limit $a\to0, N\to\infty$ for $N\,a =L$, 
we obtain the 5D gauge action in the continuum
\beq
S_5^G = \,\int d^4 x \int^L_0 dy \sqrt{g}
\left\{-\frac{1}{2\,g_5^2}{\rm Tr}[F^2_{MN}]\right\}~.
\label{fdgaugeaction}
\eeq

%---------------------------------------------------------------------------
\subsection{Deconstruction of the Warped Fermion Theory} 
\label{sec:fermiondeconstruct}
\begin{figure}[ht!]
\vspace{0.3cm}
\centerline{ \hspace*{0.0cm}
\includegraphics[width=0.85\textwidth]{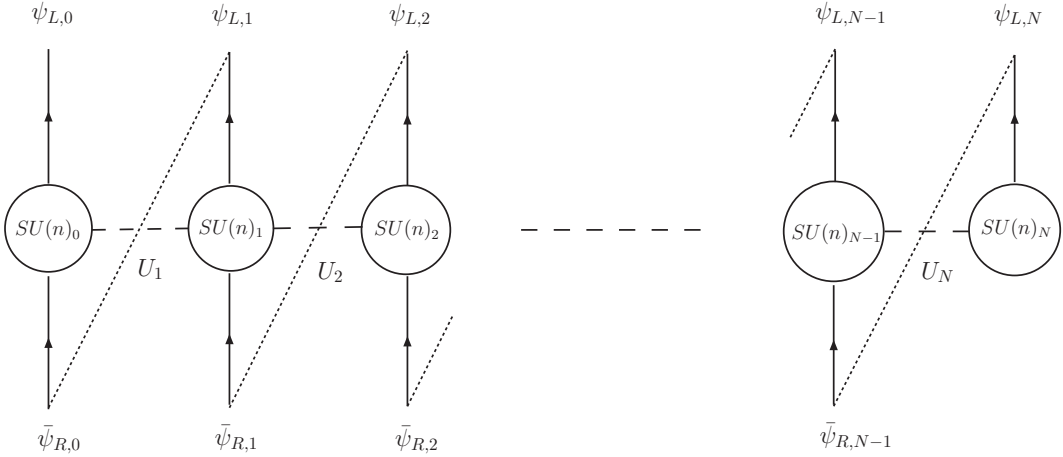}
}
\caption{\em Moose diagram to deconstruct the warped extra dimension model with fermions. The circles represent $SU(n)$ gauge groups. The zeroth and $N$-th sites are identified 
as the UV and IR branes in the continuum theory, respectively. 
The out-going (in-going) arrows represent chiral fermions in the fundamental (anti-fundamental) representation of $SU(n)$.
The link scalar fields, $U_i$, are 
$(n, \bar{n})$ under $SU(n)_{i-1}\times SU(n)_i$.  
The dotted lines represent Yukawa couplings for chiral fermions.
Boundary conditions imply the absence of $\bar\psi_{R,N}$ leading to a left-handed zero mode.
To obtain a right-handed zero mode, one has a similar moose diagram with the same hopping direction but different boundary conditions, which correspond to removing $\psi_{L,0}$. 
}
\label{fig:moose}
\end{figure}

%----------------------------------------------------------------------------
In order to write down the deconstructed version of the fermion theory in warped extra dimensions, it is  convenient to rewrite the fermion action in (\ref{actionall}) as
\beq
S_5^f = \,\int d^4 x \int_0^Ldy\,\left\{ e^{-3ky}\,\bar\Psi i\gamma_\mu D^\mu\Psi + e^{-4ky}\,M_\Psi\bar\Psi\Psi
-\,e^{-4ky}\,\bar\Psi\gamma_5\,\overleftrightarrow{\partial_5} \Psi
\right\}
\eeq
in the $A_5=0$ gauge,  
with $\overleftrightarrow{\partial_5}\equiv (1/2)(\overrightarrow{\partial_5}-\overleftarrow{\partial_5})$. 
Naively deconstructing this 5D theory results in the $N+1$ site action
\beqa
S_5^f &= &\,\int d^4 x \sum_{j=0}^N\,\left\{ 
\bar\psi_L^ji\slsh\partial\psi_L^j + \bar\psi_R^ji\slsh\partial\psi_R^j
+e^{-kaj}\,M_{\Psi}\,\bar\psi^j\psi^j \right.\nonumber\\
&&\left.+ \frac{e^{-kaj}}{2a}\,\left(\bar\psi^j_R\psi^{j+1}_L - \bar\psi^j_L\psi^{j+1}_R + {\rm h.c.}\right)
\right\}\,,
\label{beforewilson}
\eeqa
which is obtained after proper normalization of the fermion kinetic terms (absorbing $e^{-3ky/2}$ into the fermion field). However, the theory described by (\ref{beforewilson})
is not the correct discretization of the continuum action since it leads to doubling of all 
levels, and in particular to 
two massless chiral fermions, i.e. two zero modes. 
This is a reflection of the well known fermion doubling problem in lattice gauge theories. A solution to this problem is the introduction
of a Wilson term in the 5D action~\cite{wilsonterm} of the form 
\beq
S_W = \,\eta\,a\,\int d^4 x\,\int_0^L dy \sqrt{g}\,\bar\Psi\left(\partial_5 \right)^2\Psi~,
\label{waction}
\eeq
where at this point $\eta$ is an arbitrary coefficient. The Wilson term in (\ref{waction}) is a higher-dimensional operator 
suppressed by $a$, and therefore vanishes in the continuum limit. The discretization of 
the compact dimension in (\ref{waction}) gives
\beq
S_W = \,\eta\,\int d^4 x \, \sum_{j=0}^N\,\frac{e^{-kaj}}{a}\,\left\{
\bar\psi^j_L\psi^{j+1}_R + \bar\psi^j_R\psi^{j+1}_L -2\bar\psi^j_L\psi^j_R + {\rm h.c.}
\right\}~,
\label{discretewaction}
\eeq
where we have already properly normalized fermion fields. The full discretized action is then obtained when adding 
(\ref{discretewaction}) to (\ref{beforewilson})
\beqa
S_5^f + S_W&= &\,\int d^4 x \sum_{j=0}^N\,\left\{ 
\bar\psi_L^ji\slsh\partial\psi_L^j + \bar\psi_R^ji\slsh\partial\psi_R^j
+e^{-kaj}\,(M_{\Psi}-\frac{2\eta}{a}\,)\,\bar\psi^j\psi^j \right.\nonumber\\
&&\left.+ \left[ (\eta-\frac{1}{2})\,\frac{e^{-kaj}}{a}\bar\psi^j_L\psi^{j+1}_R 
+(\eta+\frac{1}{2})\,\frac{e^{-kaj}}{a}\bar\psi^j_R\psi^{j+1}_L + {\rm h.c.}\right]
\right\}~.
\label{afterwilson}
\eeqa
We can see that by choosing $\eta=\pm 1/2$ it is possible to eliminate one of the hopping 
directions in the lattice, which results in removing one of the two zero modes. For instance, for $\eta=1/2$, 
we obtain
\beqa
 S_5^f + S_W&= &\,\int d^4 x \sum_{j=0}^N\,\left\{ 
\bar\psi_L^ji\slsh\partial\psi_L^j + \bar\psi_R^ji\slsh\partial\psi_R^j
+\frac{e^{-kaj}}{a}\,(cka-1)\,\bar\psi^j\psi^j \right.\nonumber\\
&&+\left.\left(\,\frac{e^{-kaj}}{a}\bar\psi^j_R\psi^{j+1}_L + {\rm h.c.}\right)
\right\}~.
\label{etamhalf}
\eeqa
corresponding to only one hopping direction, as  illustrated in Figure~\ref{fig:moose}. 

Thus, the 5D theory can be written as a purely four-dimensional  
model corresponding to 
the  moose diagram in Figure~\ref{fig:moose}. 
The generic form of the Lagrangian of this $N+1$ site moose diagram is given by
\beqa
{\cal L}&=& -\frac{1}{2}\,\sum^N_{j=0}\,\Tr \left[F^j_{\mu\nu}\,F^{\mu\nu, j}\right] 
\,+\, \sum^N_{j=1} \Tr\,|\partial_\mu U_j + i\,g\,A^{j-1}_\mu\,U_j  
- i\,g\,U_j\,A^{j}_\mu|^2 \nonumber \\
&& + \sum^N_{j=1} \lambda\, \Tr\, ( \bar{\psi}_{R, j-1}\,U_j\,\psi_{L, j}   + h.c.)
\,+\,\sum^N_{j=0} \,\Tr\,(\mu_j\,\bar{\psi}_{L, j}\,\psi_{R, j}\,+\,h.c.)\,+\,\cdots\,,
\eeqa
where we have used the canonical kinetic terms for the gauge fields and absorbed the group generator into $A^j_\mu$, and ``$\Tr$" acts on the group indexes. 
Here, we have not included the scalar potential, which fixes the link 
field VEVs $\langle U_j \rangle = \frac{v_j}{\sqrt{2}}=\frac{v}{\sqrt{2}}\,q^j$. In order to match to the theory in the continuum limit, we have found that the following conditions should be satisfied, in addition to the dictionary in Eq.~(\ref{eq:dictionary}):
\beqa
\mu_j = -g\,v\,q^{c +  j - 1/2}\,, \qquad \lambda = \sqrt{2}\,g\,,\qquad q \rightarrow 1_-\,.
\label{eq:conditions}
\eeqa
Just as for the case of gauge bosons, here $q<1$ and $q \rightarrow 1_-$ corresponds 
to taking the continuum limit. The $c$  parameter in the matching condition 
(\ref{eq:conditions}) for $\mu_j$ will be identified as 
the bulk mass parameter in the continuum theory, which controls the localization of 
the fermion zero mode.

We can repeat the same procedure followed for the gauge bosons in order to 
obtain the difference equations leading to the solutions for the spectrum of fermion modes.
The fermion mass-squared matrix in the basis $(\psi_{L, 0}, \psi_{L, 1}, 
\cdots, \psi_{L, N})$ can be written as:
\beqa
m^T\,m\,=\,\left(  
\renewcommand{\arraystretch}{1.6}
\begin{array}{ccccccc}
\mu_0^2   & g\,\mu_0\,v_1   & 0               &   0   & \cdots    & 0   &    0     \\
g\,\mu_0\,v_1  &  g^2v_1^2\,+\,\mu^2_1  &  g\,\mu_1\,v_2     &  0  & \cdots  &   0    &0      \\
0    &   g\,\mu_1\,v_2     &  g^2v_2^2 + \mu^2_2   &  g\,\mu_2\,v_3  & \cdots & 0 & 0   \\
\vdots & \vdots & \vdots  & \vdots &  \cdots  & \vdots& \vdots   \\
0    & 0 & 0 &0 &\cdots &   g^2v^2_{N-1}\,+\,\mu^2_{N-1}     &  g\,\mu_{N-1}\,v_N\,  \\
0 & 0 & 0&0 &\cdots &        g\,\mu_{N-1}\,v_N   & \mu_N^2
\end{array}
\right)\,.
\eeqa
Using the orthonormal rotation matrix $\psi_{L, j} =\sum^N_{n=0} h^L_{j, n}\,{\cal \psi}_{L, (n)}$ 
and substituting $\mu_j$ in Eq.~(\ref{eq:conditions}) into the above equation, 
we arrive at the following difference equations
\beqa
\left(q^{-(c+\frac{1}{2})}\,+\,q^{(c+\frac{1}{2})}\,-\,q^{-(c+\frac{1}{2})} 
(x_n\,q^{-j})^2\right)\,h^L_{j, n}\,-\,q\,h^L_{j+1, n}\,
-\,q^{-1}\,h^L_{j-1, n}\,=\,0\,.
\label{eq:fermionLEOM}
\eeqa
Similarly, for the right-handed fermions, we obtain
\beqa
\left(q^{-(c-\frac{1}{2})}\,+\,q^{(c-\frac{1}{2})}\,-\,q^{-(c-\frac{1}{2})} (x_n\,q^{-j})^2\right)\,h^R_{j, n}\,
-\,q\,h^R_{j+1, n}\,-\,q^{-1}\,h^R_{j-1, n}\,=\,0\,.
\eeqa
The solutions to the above two difference equations are
\beqa
h^{L, R}_{j, n}\,=\,R^{L, R}_n\,q^{-j}\,\left[ J_{|c\pm\frac{1}{2}|}(x_n\,q^{-j}; q^2)\,
+\,b_{|c\pm\frac{1}{2}|}(x_n; q^2)\,Y_{|c\pm\frac{1}{2}|}(x_n\,q^{-j}; q^2)               
\right]\,,
\eeqa
with the ``+" sign for left-handed fermions and ``--" sign for right-handed fermions, and where  $R^{L, R}_n$ are normalization factors.  These solutions match, in the continuum limit,  to the general solutions 
of the 5D theory given in (\ref{fermisol}). 
To obtain a chiral zero mode, we choose the boundary condition $h^R_{N, n} = 0$ for all  $n$ 
to get a left-handed zero-mode fermion, and $h^L_{0, n} = 0$ 
for all $n$ to get a right-handed zero mode. 
These boundary conditions are equivalent to removing $\psi_{R, N}$ or $\psi_{L, 0}$ 
from the theory and are illustrated in Fig.~\ref{fig:moose}. For instance, for the case of a left-handed zero mode and solving Eq.~(\ref{eq:fermionLEOM}), we have 
\beq
\frac{h^L_{j+1, 0}}{h^L_{j, 0}} = q^{c_L\,-\,\frac{1}{2}}\,.
\eeq
Since $q<1$, 
the left-handed zero mode is therefore ``localized'' in theory space toward the left side of the moose diagram for $c_L > 1/2$, 
whereas for $c_L<1/2$,  toward the $N$-th site. 
Then, upon taking the continuum limit this choice matches the corresponding behavior of a left-handed zero mode in the continuum theory, by 
identifying the zeroth site with the UV brane and the $N$-th site with the IR brane.
Conversely, for the right-handed zero mode we obtain:
\beq
\frac{h^R_{j+1, 0}}{h^R_{j, 0}} = q^{-(c_R\,+\,\frac{1}{2})}\,,
\eeq
so that for $c_R>-1/2$ the right-handed zero mode is ``$N$-th-site'' (IR) localized, whereas for $c_R<-1/2$ it is localized towards the zeroth site corresponding to UV localization in the continuum.

It is useful to consider the behavior of these solutions in various limits and compare them to the continuum limit case. 
For instance, if we are considering a left-handed zero mode solution in the $\mu_j/v_j\to 0$ limit, we see that this requires  $c_L\gg 1/2$, which corresponds to extreme UV localization in the continuum theory. On the other hand, the limit 
$\mu_j/v_j\to\infty$ requires $c_L\ll 1/2$, which corresponds to extreme IR localization in the continuum.
Conversely, for a right-handed zero mode the limit $\mu_j/v_j\to 0$ leads to $c_R\gg -1/2$, corresponding to IR localization  in the continuum, with the limit $\mu_j/v_j\to\infty$ corresponding to a UV-localized right-handed zero mode.  The matching between the continuum theory and the discretized theory for various limits is illustrated in Table~\ref{tab:extremlimit}.
\begin{table}[htdp]
\renewcommand{\arraystretch}{1.8}
\begin{center}
\begin{tabular}{ccc}
\hline
left-handed fermion & \hspace{1cm}   &right-handed fermion  \\
$c_L \gg \frac{1}{2} \; (\mbox{UV}) \quad \leftrightarrow \quad \frac{\mu_j}{v_j} \rightarrow 0$ 
&  \hspace{1cm} &  $c_R \gg - \frac{1}{2} \; (\mbox{IR}) \quad \leftrightarrow \quad \frac{\mu_j}{v_j} \rightarrow 0$ \\ 
$c_L \ll \frac{1}{2} \; (\mbox{IR}) \quad \leftrightarrow \quad \frac{\mu_j}{v_j} \rightarrow \infty$ 
&  \hspace{1cm}  & $c_R \ll - \frac{1}{2} \; (\mbox{UV}) \quad \leftrightarrow \quad \frac{\mu_j}{v_j} \rightarrow \infty$ \\ \hline 
\end{tabular}
\end{center}
\caption{\em Matching of the continuum theory and the discretized theory for different limits.}
\label{tab:extremlimit}
\end{table}%
In the next section we will make use of results obtained in these limits in order to compute the 
low energy interactions induced after requiring anomaly cancellation.

%----------------------------------------------------------
\section{Anomaly Cancellation}
%From Wess-Zumino-Witten to Chern-Simons Terms
%----------------------------------------------------------
\label{wzwtocs}
Having completed our understanding of the deconstructed version of warped extra dimensional theories and their 
continuum limit, we are now in a position to study the necessary ingredients for anomaly cancellation  in these theories. 
We will derive the Chern-Simons terms for different gauge theories in warped extra dimensions, starting from the deconstructed theory. 
However, one should notice that the procedure described in this section can also be applied to flat extra-dimensions,  
since the CS terms only depend on the  topological properties of gauge theories, and therefore should be independent of a  particular 
geometry. As an example, we explicitly work out the simplest case with a $U(1)$ gauge group propagating 
in the bulk. We first calculate the WZW terms based on the moose diagram in the 4D theory, and then take the continuum
limit to obtain the 5D CS terms. We then move to compute the CS terms for non-Abelian as well as product gauge groups. 

For the case of a $U(1)$ gauge group, we consider two ``bulk'' fermions, $\Psi$ and $X$, which have $\psi_L^{(0)}$  and $\chi_R^{(0)}$ 
as their left-handed and right-handed zero modes, respectively. Under the $U(1)$  gauge group, $\Psi$ and $X$ have the same charge $Q$, 
so the 4D anomaly is canceled for the unbroken $U(1)$ gauge group in the low energy theory. However, in the 5D theory or in its deconstructed version, 
the anomaly is not canceled without additional terms. This can be seen from Fig.~\ref{fig:moose}, where there is a triangular anomaly 
at the $N$-th site for the $\Psi$  field. Similarly, there is a triangular anomaly on the zeroth site for the $X$ field.  
These triangle anomalies can be canceled in the continuum theory by adding an appropriate CS term~\cite{Hill:2006ei,ArkaniHamed:2001is}.  
Here, we also want to show how the anomaly cancellation works in the deconstructed theory and how to match to the CS term in the continuum. 

Preserving the gauge symmetry, the deconstructed Lagrangian is 
\beqa
{\cal L}&=& -\frac{1}{4}\,\sum^N_{j=0}\,F^j_{\mu\nu}\,F^{\mu\nu, j}
\,+\, \sum^N_{j=1} \,|\partial_\mu U_j + i\,g\,A^{j-1}_\mu\,U_j  
- i\,g\,U_j\,A^{j}_\mu|^2 \nonumber \\
&& + \sum^N_{j=1} \lambda\, ( \bar{\psi}_{R, j-1}\,U^Q_j\,\psi_{L, j}   + h.c.)
\,+\,\sum^N_{j=0} \,(\mu_j\,\bar{\psi}_{L, j}\,\psi_{R, j}\,+\,h.c.)\,+\,\cdots\,,
\eeqa
The link field $U_j$ is charged as $(1, -1)$ under $U(1)_{j-1}\times U(1)_j$. Integrating out heavy fermions with chiral masses, results in the appearance of WZW terms  
in the low energy theory. In the case at hand, to integrate out all fermions other than the zero-mode and for the most generic case  $\mu_j/v_j\sim O(1)$, the action obtained  will depend on the localization of $\psi^{(0)}_L$ and $\chi^{(0)}_R$ through these ratios~\cite{paper2}. Although it is  possible  to obtain the WZW terms generated, their sum in the continuum limit has a non-trivial dependence  
on $c_R$ and $c_L$ through non-local terms in the extra dimension. 
On the other hand, it is quite simple to obtain the continuum result for the cases with $\mu_j/v_j=0$ and 
$\mu_j/v_j\to\infty$.

\begin{figure}[ht!]
\vspace{0.3cm}
\centerline{ \hspace*{0.0cm}
\includegraphics[width=0.7\textwidth]{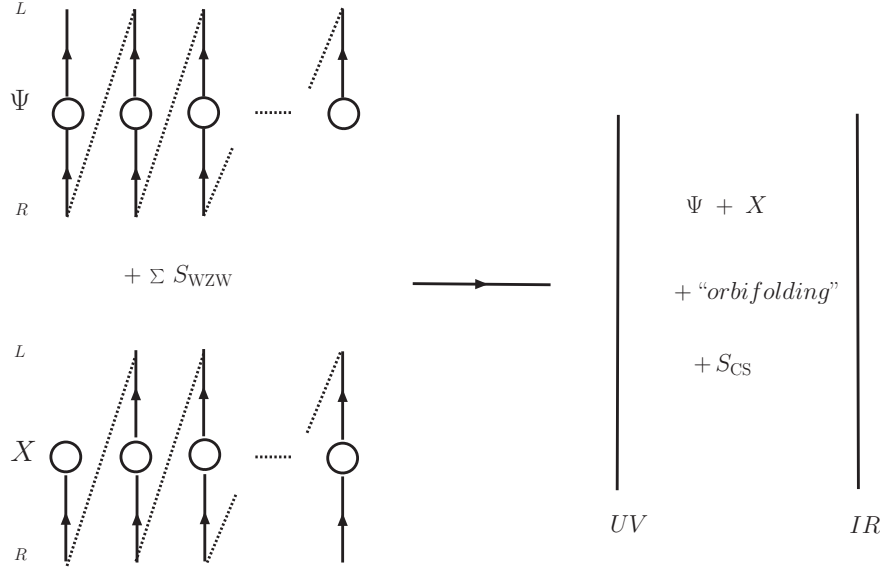}
}
\caption{\em The anomaly cancellation for two fermions $\Psi$ and $X$ propagating in the bulk. 
The ``orbifolding" is chosen to have a left-handed zero mode for $\Psi$ and a right-handed zero mode for $X$. 
To cancel  the gauge anomalies, a summation of WZW terms is needed in the deconstructed theory, corresponding to a  
CS term in the continuum theory. 
}
\label{fig:anomalycancel}
\end{figure}

First, let us consider the $\mu_j/v_j=0$ limit for both $\Psi$ and $X$ fermions. As discussed at the end of the previous section and in Table~\ref{tab:extremlimit}, this corresponds to   an extremely UV-localized $\psi^{(0)}_L$ and   an extremely IR-localized $\chi^{(0)}_R$. For the $\Psi$ field,   
the fermion mass matrix is ``diagonal" in the sense that $\psi_{R, j-1}$ 
and $\psi_{L, j}$ form  a massive Dirac fermion without mixing with other fermions. So, integrating out the massive fermions 
$\psi_{R, j-1}$ and $\psi_{L, j}$, we arrive at a summation of WZW terms~\cite{Manohar:1984uq}
\beqa
{\cal S}_{\rm eff}&=&\sum_{j=1}^N\,{\cal S}_{\rm WZW}(A_{j-1}, A_{j}, U_j)\nonumber \\
&=&\frac{1}{48\pi^2}\,\int\,\sum_{j=1}^N\,\left[\alpha_4(QA_j,\, \xi_j^Q\,d\,\xi_j^{Q\dagger})\,-\,
\alpha_4(QA_{j-1},\, \xi_j^{Q\dagger}\,d\,\xi_j^{Q})\,-\, {\cal B}(A_{j-1}^{\xi_j^{Q}}, A_{j}^{\xi_j^{Q\dagger}} )   \right]\,.
\eeqa
Here, we defined $\xi_j$ by $U_j\equiv \xi_j^2$; $A_{j-1}^{\xi_j^{Q}}\equiv QA_{j-1}+\xi_j^{Q\dagger}\,d\,\xi^Q_j$, and  
the Bardeen counter-term is defined by ${\cal B}(A_1, A_2) = 2(dA_1 + dA_2)A_1A_2$ for this case at hand. 
We make use of 1-form notation such that $A\equiv g\,A_\mu\,d\,x^\mu$ and $d\equiv dx^\mu\partial/\partial x^\mu$. We omit the 4D
Levi-Civita $\epsilon$ tensor, so that any product of  1-forms and their derivatives are contracted by it.   
The 4-form $\alpha_4$ can be calculated by acting the homotopy operator on the CS 5-form and has the expression  
$\alpha_4(A, B)\equiv 2\,dA\,A\,B$  for the $U(1)$ case~\cite{Zumino:1983ew}. After some algebraic manipulations, we obtain
\beqa
{\cal S}_{\rm eff}=\frac{Q^3}{48\,\pi^2}\,\int\,\sum_{j=1}^N\,\left[
2\,A_{j-1}\,d\,A_{j-1}\,A_j\,+\,2\,A_{j-1}\,d\,A_{j-1}\,U_j\,d\,U_j^\dagger\,+\,U^\dagger_j\,d\,U_j\,d\,A_{j-1}\,A_j\,-\,p.c.
\right]\,,
\eeqa
where $p.c.$ denotes parity conjugation such that $A_{j-1} \leftrightarrow A_j$ and $U_j \leftrightarrow U_j^\dagger$ (for the non-Abelian case, see Ref.~\cite{Skiba:2002nx}).
Taking the continuum limit ($a\rightarrow 0$), we identify $A_{j-1}= A(y)$, $A_{j}=A + a\,\partial_5 A$ and $U_j = 1 + a A_5$, 
and keep only the terms of order $a$. Identifying $a\Sigma_j$  as $\int dy$, we arrive at the CS term in the 5D theory:
\beq
{\cal S}_{\rm eff}=-{\cal S}_{\rm CS}\,=\,\frac{-\,Q^3}{24\,\pi^2}\int A\,d\,A\,d\,A \,=\,\frac{-\,Q^3}{24\,\pi^2}\,\int d^5x\,
\epsilon^{ABCDE}A_A\,\partial_B\,A_C\,\partial_D\,A_E\,.
\label{minus_cs}
\eeq
One can check that all anomaly terms can be canceled by adding the CS term into the 5D continuous theory~\cite{Hill:2006ei}. 
It is easier to understand this cancellation  in the deconstructed moose theory. 
For the $\Psi$ field, the boundary conditions are equivalent to removing 
the right-handed particle $\psi_{R, N}$ at the last site to obtain a left-handed zero mode. 
Therefore, there exists a triangle anomaly at the last site by this ``orbifolding" procedure. 
For the $X$ field with a right-handed zero mode, its boundary conditions are equivalent to removing the left-handed fermion at the zeroth site. 
Altogether, we have the ``brane" localized triangle anomalies given by
\beqa
\delta S_{\rm branes}=\frac{Q^3}{24\,\pi^2}\,\int\,\theta_{N}\,d\,A_N\,d\,A_N\,-\, \theta_{0}\,d\,A_0\,d\,A_0 \,,
\eeqa
with $\theta_j\equiv \theta_j(x)$ the gauge transformation parameter for each site. Here, $\delta S$ is the variation of the action under gauge transformation. Performing gauge transformations on the CS terms, we have
\beqa
\delta S_{\rm CS}= \frac{Q^3}{24\,\pi^2}\,\int\,\sum^N_{j=1}
\left[ \theta_{j-1}\,d\,A_{j-1}\,d\,A_{j-1} \,-\,   \theta_{j}\,d\,A_{j}\,d\,A_{j} \right]\,=\,-\delta S_{\rm branes}\,.
\eeqa
Indeed, the addition of the variation of the CS term and the brane-localized triangle anomalies cancel, making the full theory anomaly free.  The anomaly cancellation in both 
the deconstructed and continuum theories is depicted in the Fig.~\ref{fig:anomalycancel}.
The top figure in the left panel in Fig.~\ref{fig:anomalycancel} depicts the deconstructed theory with a left-handed zero mode $\psi^{(0)}_L$, with the one at the 
bottom showing the case of a right-handed zero-mode $\chi^{(0)}_R$. To cancel the chiral anomalies on the end sites of the moose diagram, 
we add a summation of WZW terms. In the continuum limit, shown  in the right panel of Fig.~\ref{fig:anomalycancel}, 
the WZW terms lead to the CS term and the full theory is anomaly free.   

We close this section by generalizing the 
procedure described above for the Abelian case, to derive the CS terms for non-Abelian and product gauge groups, which 
will be used later. For the non-Abelian case the CS terms are
\beq
S_{\rm CS} = \frac{1}{24\,\pi^2}\int \Tr\left[ A\,d\,A\,d\,A\,+\,\frac{3}{2}\,A^3\,d\,A\,+\,\frac{3}{5}\,A^5 \right], 
\label{eq:CSnonabelian}
\eeq
which is non-zero only if the group has a non-zero fully-symmetric structure constant or equivalently 
$d^{abc}=\Tr[t^a\{t^b, t^c\}]\ne 0$. So, for $SU(2)$, there is no second Chern-Simons character and no corresponding terms in Eq.~(\ref{eq:CSnonabelian}). 

To obtain a complete set of CS terms for realistic models, 
we also need to obtain the WZW terms associated with product of gauge groups. For example, 
let us consider the SM gauge bosons propagating in the bulk of the extra dimension and one bulk fermion with charges $(3, 2, Y)$ 
under $SU(3)_c\times SU(2)_W \times U(1)_Y$. All the CS terms can be obtained simply by replacing $A$ in Eq.~(\ref{eq:CSnonabelian}) 
by $A= G+W+ YB$. Here, $G$, $W$ and $B$ are the gauge boson fields of $SU(3)_c$, $SU(2)_W$, and $U(1)_Y$ in the one-form. 
The trace in Eq.~(\ref{eq:CSnonabelian}) is replaced by $\Tr= \Tr_3\,\Tr_2\,\Tr_1$ with the $\Tr_i$'s acting on different gauge space 
and $\Tr_1=1$. So, we have the CS terms for a product of gauge groups given by 
\beqa
S_{\rm CS} &=& \frac{1}{24\,\pi^2}\int {\Tr_3\,\Tr_2\,\Tr_1}\left[ (G+W+ YB)\,d\,(G+W+ YB)\,d\,(G+W+ YB) \right.\nonumber \\ 
&&\left.\hspace{1.5cm}\,+\,\frac{3}{2}\,(G+W+ YB)^3\,d\,(G+W+ YB)\,+\,\frac{3}{5}\,(G+W+ YB)^5 \right],   \nonumber \\
&=&  \frac{1}{24\,\pi^2}\int N_c N_w Y^3 B\,d\,B\,d\,B\,+\,N_w\,\Tr_3\,\left[ G\,d\,G\,d\,G\,+\,\frac{3}{2}\,G^3\,d\,G\,+\,\frac{3}{5}\,G^5         
\right] \nonumber \\
&& \hspace{1.5cm}+\,3\,N_c\,Y\,B\,\Tr_2\,\left[  (dW + W^2)^2\right]\,+\,3\,N_w\,Y\,B\,\Tr_3\,\left[  (dG + G^2)^2\right] \nonumber \\
&& \hspace{1cm}\,+\,{\rm boundary\; terms}\,,
\label{csinproduct}
\eeqa
with total derivative terms neglected and $N_c=3$ and $N_w=2$.
%---------------------------------------------------
\section{Integrating out Fermion KK modes}
%---------------------------------------------------
\label{ioutfermions}
Having understood  the anomaly cancellation in the extra dimension theory, we now consider the low energy theory by integrating out the fermion KK-modes, 
and study the remaining possible topological interactions among gauge bosons. 
Although anomaly cancellation is independent of the fermion localization, the topological interactions of the gauge boson KK modes indeed 
depend  on the fermion profiles in the fifth dimension. To simplify our discussions in this paper, we continue working in 
the limits with $\mu_j/v_j=0$ or $\mu_j/v_j\to\infty$ in the deconstructed theory, equivalent to fermion zero-modes extremely localized on the IR or UV 
branes in the continuum. We believe that these limits capture the general features of warped extra dimension models, 
where the all fermion zero modes are localized close to the UV brane except for the right-handed top quark, which typically is highly localized close to the IR brane.

\begin{figure}[ht!]
\vspace{0.3cm}
\centerline{ \hspace*{0.0cm}
\includegraphics[width=0.7\textwidth]{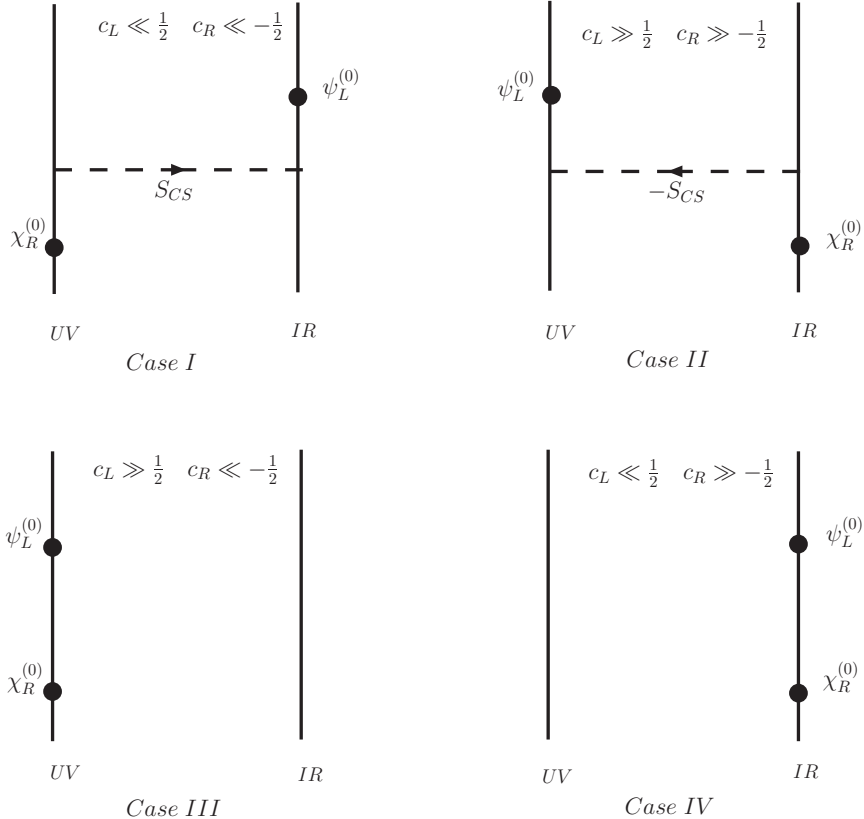}
}
\caption{\em Four different cases for the remaining low energy theory after integrating out heavy fermions. All cases are anomaly free. For the Case I and II, the left-handed and the right-handed zero modes are localized in different branes. This leads to a CS term, which contains physical interactions among gauge boson KK-modes. For the Case III and IV, the left-handed and the right-handed zero modes are localized in the same brane and no physical topological interactions are left. 
}
\label{fig:locfermions}
\end{figure}

For the simple $U(1)$ example in 
Fig.~\ref{fig:anomalycancel}, we use the deconstructed theory as a guide to perform the integration 
of the fermion KK modes. Taking the limit $\mu_j/v_j\to\infty$ for both $\Psi$ and $X$, there are effectively 
no chiral-symmetry breaking links and all we have are vector-like fermions, except on the end sites of the moose diagram. Then, after we integrate out 
these heavy vector-like fermions, no additional WZW terms are generated in the deconstructed theory. So, in the low energy theory,  we have 
one left-handed zero mode $\psi^{(0)}_L$ on the $N$-th site, one right-handed zero mode $\chi^{(0)}_R$ 
on the zeroth site and the original summation of WZW terms. Referring back to the continuum theory, 
we have  $\psi^{(0)}_L$ on the IR bane, $\chi^{(0)}_R$ on the UV brane and the original CS term in the bulk, 
which is schematically shown as the Case I of Fig.~\ref{fig:locfermions}. 
The CS term in the bulk contains the topological interactions among gauge bosons. 

Taking the $\mu_j/v_j\to 0$ limit for both fermions and integrating fermion KK modes, 
the left-handed zero mode $\psi_L^{(0)}$ is localized on the zeroth site and the right-handed zero mode $\chi_R^{(0)}$ 
is on the $N$-th site. Furthermore, there is a sum of WZW terms corresponding to the $\psi_L^{(0)}$ tower and another one 
corresponding to  the $\chi^{(0)}_R$ tower. 
One of the sums of WZW terms cancels the original WZW terms and leaves just one sum of WZW terms, 
which has opposite sign with respect to the original one. 
In the continuum limit, shown in the Case II of Fig.~\ref{fig:locfermions}, we have $\psi^{(0)}_L$ on the UV brane and $\chi^{(0)}_R$ on the IR brane. 
The summation of CS terms is $2\,S_{\rm eff}+ S_{\rm CS}= -\,S_{CS}$, with $S_{\rm eff}= -S_{\rm CS}$ in Eq.~(\ref{minus_cs}) corresponding 
to the summation of WZW terms. The remaining action is again anomaly free. 
For most of warped extra dimension models the right-handed zero mode of the top quark is localized close to  the IR brane, 
whereas the left-handed zero mode is moderately UV-localized.  
Therefore, the Case II in Fig.~\ref{fig:locfermions} can be used as an approximation for  the top quark contributions 
in a realistic model. 
The remnant CS term in the bulk contains physical interactions among gauge boson KK modes. This result agrees with our intuition 
in a sense that because of the different localizations of top quark left and right-handed zero modes, a nontrivial topological interaction 
remains in the low energy theory.

Let us now consider the case where both the left-handed and right-handed zero modes are UV-localized. 
In the deconstructed picture this is 
achieved by taking the $\mu_j/v_j\to 0$ limit for the $\psi_L^{(0)}$ tower, and $\mu_j/v_j\to\infty$ for the $\chi_R^{(0)}$ one. 
Integrating out fermion KK modes, there is only one summation of WZW terms generated from the $\psi_L^{(0)}$ tower, which is canceled 
by the original WZW terms in the theory. 
The $\chi_R^{(0)}$ tower has only vector-like mass terms and therefore does not lead to any WZW terms. 
Therefore, as illustrated in the Case III of Fig.~\ref{fig:locfermions}, 
we only have two chiral zero modes localized in the UV brane and no additional terms in the bulk. The same result is obtained for the case 
with both zero modes localized in the IR, by switching the limits, 
and is shown in the Case IV of Fig.~\ref{fig:locfermions}. 
In most realistic warped extra dimension models, and in order to satisfy various constraints including electroweak precision observables and 
flavor changing processes, the first two generations of SM fermions have both left and right-handed zero modes localized towards the UV brane. 
So, Case~III can be thought of as an approximate description of  the first two generations of  fermions as well as the bottom quark. 
Then, we see that they do not contribute  new physical topological interactions. 
Once again, this result agrees with the intuition that if the left and right-handed zero-mode 
fermions have the same profile in the fifth dimension, the theory is ``vector-like" and no new 
topological interactions should be generated.

Summarizing the discussion above, we see that when both chiralities of the zero modes are localized at the same 
fixed point there are no remnant interactions, whereas such interactions are generated when left and right-handed zero modes are
localized at different ends of the extra dimension. 
At least in these simplified cases, obtained in the extreme limits 
$\mu_j/v_j\to 0$ and/or $\mu_j/v_j\to\infty$, this confirms the intuition that the presence of these terms is associated with 
the different localization of left and right-handed zero modes in the bulk.  
Finally, in the more general case with finite values of the bulk fermion masses, 
we expect that the form of the remnant interactions should depend on the bulk zero-mode wave-functions, i.e. 
on the bulk mass parameters $c_L$ and $c_R$. It is possible to obtain this general dependence in the 
deconstructed description~\cite{paper2}. However, the continuum limit of the general case will have a complicated non-local 
dependence on $c_L$ and $c_R$. In order to make things more transparent, we will only 
consider the simplified limiting cases in the rest of the paper. They should give us a good estimate of the types of 
physical effects we can expect.

\section{Topological Physical Processes}
\label{topproc}
In this section we show how the CS terms described in the previous section lead to actual 
novel physical processes, as opposed to being just an artifact to cancel the anomalies. 
Once again, we step back to the deconstructed description in order to better understand the 
presence of these terms. Throughout the rest of the paper we will make use of the results obtained for the limit
$\mu_j/v_j\to 0$, corresponding to a UV-localized left-handed zero mode and an IR-localized 
right-handed zero mode (see Case II in Fig.~\ref{fig:locfermions}). 
As we discuss in the next section, we will use this setup as an approximation to warped extra dimension models where 
$t_R$ is the only fermion significantly localized towards the TeV brane.

\begin{figure}[ht!]
\vspace{0.3cm}
\centerline{ \hspace*{0.0cm}
\includegraphics[width=0.7\textwidth]{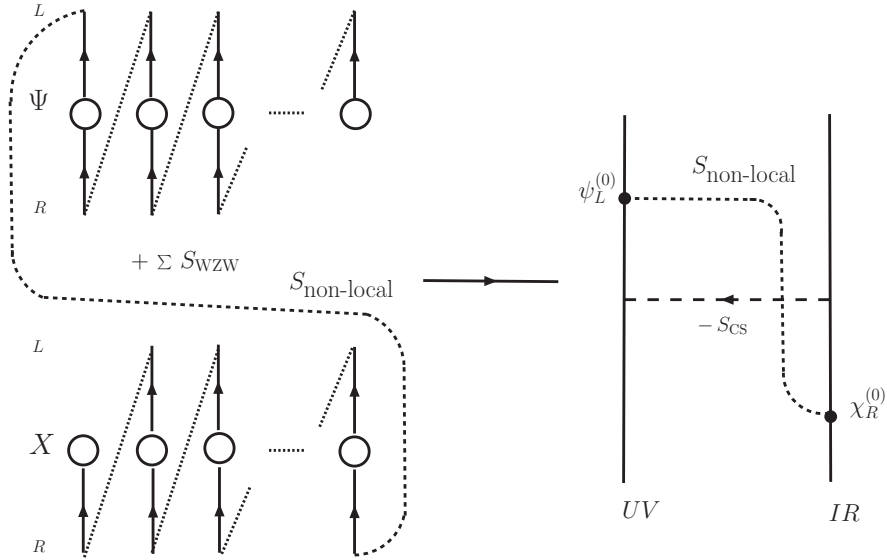}
}
\caption{\em The triangular loop contribution to the gauge boson KK mode interactions from fermion 
zero modes can be replaced by a non-local link in the deconstructed theory and a Wilson-like non-local interaction in the continuum theory.
}
\label{fig:topointeraction}
\end{figure}

We are interested in physical processes involving gauge bosons and KK gauge bosons from topological interactions. 
For these processes, the localized zero-mode fermions  also contribute to gauge boson interactions 
through triangular diagrams~\footnote{Actually, those triangular loop contributions from zero-mode fermions 
are important to provide 4D gauge invariant interactions after combined with interactions from the CS terms.}. 
In the deconstruction language, their contributions can be obtained by adding a Wilson mass term for the chiral fermions: 
$\lambda\,\bar{\chi}_{R, N}\,U\,\psi_{L, 0}\,+\,h.c.$, with $U\equiv U_NU_{N-1}\,\cdots\,U_2U_1$ connecting the two end sites. 
The corresponding moose diagram and the related continuum theory are illustrated in Fig.~\ref{fig:topointeraction}.
Integrating out these two chiral fermions, 
leads to one more term in the topological Lagrangian in addition to the CS term. The total topological interactions for the $U(1)$ case now are
\beqa
{\cal S}_{\rm topo}&=&-\frac{Q^3}{48\,\pi^2}\,\int\,\sum_{j=1}^N\,\left[
2\,A_{j-1}\,d\,A_{j-1}\,A_j\,+\,2\,A_{j-1}\,d\,A_{j-1}\,U_j\,d\,U_j^\dagger\,+\,U^\dagger_j\,d\,U_j\,d\,A_{j-1}\,A_j    \right.     \nonumber \\
&& \left.  \hspace{2cm} \,+\,2\,A_{N}\,d\,A_{N}\,A_{0}\,+\,2\,A_{N}\,d\,A_{N}\,U\,d\,U^\dagger\,+\,
U^\dagger\,d\,U\,d\,A_N\,A_{0}\,-\,p.c. 
\right]\,,
\label{eq:topoDec}
\eeqa
The above interaction is gauge invariant and captures all the necessary topological interactions. 

Before proceeding, we want to emphasize the fact that 
{\em there are no topological interactions among gauge bosons in the two-site model}. The above equation is identically zero for $N=1$. 
However, for  $N=2$ in the three-site model with $(+, +)$ boundary conditions for the gauge fields 
(i.e. if the gauge symmetry is preserved in the low energy 
theory), one does have a remnant physical interaction: 
$B^2\,B^1\,d\,B^0$, with $B^i$ as the $i$'th KK-mode gauge bosons. Thus, these remnant interactions 
will also be present in the continuum, although for the 
$(+,+)$ boundary conditions they will have to involve the first and second KK modes. On the other hand, if $(+, -)$ 
boundary conditions are imposed, the zero mode 
becomes massive and two more interactions are allowed for this case: $B^2\,B^1\,d\,B^1$ and $B^1\,B^0\,d\,B^0$.

Taking the  continuum limit of  (\ref{eq:topoDec}), the product of link fields becomes a Wilson line connecting from the 
UV brane to the IR brane: 
\beq
U\,=\,{\rm exp}\left(-i\,\int^L_0\,dy\,A_5\,(y) \right)\,.
\eeq
Thus, the complete topological interactions are not just given by the CS terms, but we must also add the non-local terms 
resulting from the second line in  (\ref{eq:topoDec}). This results in 
\beqa
-S_{topo}&=&S_{CS}\,-\,\frac{Q^3}{48\,\pi^2}\int\,d^4x\,\left[ 2\,A(L)\,d\,A(L)\,A(0)\,+\, 2\,A(L)\,d\,A(L)\,U\,d\,U^\dagger   
\right. \nonumber \\ 
&&\left.   \hspace{4cm}\,+\,U^\dagger\,d\,U\,d\,A(L)\,A(0)\,-\,p.c.
\right]\,.
\label{eq:master}
\eeqa
One can explicitly check the gauge invariance of the above expression. 
The last term corresponds to the non-local link in Figure~\ref{fig:topointeraction}.  
In the unitary gauge, $A_5=0$ and $U=1$, we have
\beqa
-S_{topo}=\frac{Q^3}{24\,\pi^2}\,\int\,dx^5\,A\,d\,A\,d\,A\,+\,\frac{Q^3}{24\,\pi^2}\,\int\,
d^4x
\left[A(0)\,d\,A(0)\,A(L) \,-\,A(L)\,d\,A(L)\,A(0)
\right]  \,.
\label{eq:topointeraction}
\eeqa
This action leads to interactions among KK gauge bosons. 
In the rest of this Section, we will compute the form of certain triple and quartic 
topological interactions in various examples, which are going to be useful for phenomenological applications. 
\vskip0.5cm

\noindent{\bf Abelian Gauge Group}:

Using Eq.~(\ref{eq:topointeraction}) and decomposing the 5D gauge boson into 4D KK modes and concentrating on the zero and first KK modes, 
we have the following interactions:
\beqa
-S_{topo}&=&\frac{Q^3\,\bar{g}_1^3}{24\pi^2\,L^{3/2}}\int d^4x A^{(0)} dA^{(0)} A^{(1)}
\left\{\int^L_0 dy\,2 \left[f^1(y)\,f^0(y)\,\partial_y\,f^0(y)\,-\,f^0(y)\,f^0(y)\,\partial_y\,f^1(y)
\right]   \right. \nonumber \\
&& \left. \hspace{-1cm}\,+\,f^1(L)\,f^0(0)\,f^0(0)\,-\,f^1(0)\,f^0(0)\,f^0(L)\,-\,f^1(0)\,f^0(L)\,f^0(L)\,+\,f^1(L)\,f^0(L)\,f^0(0)
\right\}\,.
\eeqa
Here, $\epsilon^{\mu\nu\rho5\sigma}=-\epsilon^{\mu\nu\rho\sigma}$ is used, and $\bar{g}_1$ is a 5D gauge coupling with mass dimension $-1/2$.

For $(+, +)$ boundary conditions for the gauge bosons, we have a constant $f^0$. Then, it is straightforward to 
show that the coefficient of $A^{(0)}\,d\,A^{(0)}\,A^{(1)}$ vanishes.
\beqa
-S_{topo}&=&\frac{Q^3}{24\,\pi^2}\,\int d^4x\,A^{(0)}\,dA^{(0)}\,A^{(1)}\,\times\, 0\,. \qquad (+, +)
\label{eq:abelianpp}
\eeqa
This reflects the fact that the gauge symmetry is unbroken and this gauge-symmetry-violating operator should be vanishing. 
However, if there was a boundary-localized Higgs field breaking the gauge symmetry, this term will survive, with its coefficient 
suppressed by the square of the ratio of the localized VEV over the IR  brane scale.
 
For the case of $(+, -)$ boundary conditions, we have $f^i(L)=0$. For this case, only CS terms contribute to the topological interaction and the coefficient 
of $A^{(0)}\,dA^{(0)}\,A^{(1)}$ is non-zero and is 
\beqa
-S_{topo}&\approx&\frac{Q^3\,\bar{g}^3_1}{24\,\pi^2\,L^{3/2}}\,\int d^4x\,A^{(0)}\,dA^{(0)}\,A^{(1)}\, (-\,2.4\,\sqrt{kL}) \nonumber \\
&\approx& \frac{Q^3\,g^3_1}{24\,\pi^2}\,\int d^4x\,A^{(0)}\,dA^{(0)}\,A^{(1)}\, (-\,2.4\,\sqrt{kL})  \qquad (+, -) \,,
\label{eq:abelianpm}
\eeqa
with the 4D gauge coupling $g_1\equiv \bar{g}_1/\sqrt{L}$ and $kL \gg 1$.   
Once again, the non-zero value of the coefficient reflects the fact that all gauge symmetries are broken.
\vskip0.5cm

\noindent{\bf Non-Abelian Gauge Group} 

We consider the non-Abelian gauge group $SU(3)_c$.  In the unitary gauge, the total 
topological interaction is given by
\beqa
-S_{topo}&=& \frac{1}{24\,\pi^2}\int \Tr\left[ G\,d\,G\,d\,G\,+\,\frac{3}{2}\,G^3\,d\,G\,+\,
\frac{3}{5}\,G^5 \right]  
\,-\,\frac{1}{48 \pi^2}\,\int\,\Tr\,\Bigl[ G(L)\,d\,G(L)\,G(0) \Bigr.  \nonumber \\
&& \Bigl. \hspace{1cm}  \,+\,d\,G(L)\,G(L)\,G(0)\, - \,G(0)\,d\,G(0)\,G(L) \, -\, d\,G(0) 
\,G(0)\,G(L) 
\Bigr. \nonumber \\
&&\Bigl. \hspace{1cm} \,+\, G^3(L)\,G(0) \,-\, G^3(0)\,G(L)\, -\,\frac{1}{2}\,G(0)\,G(L)
\,G(0)\,G(L)
\Bigr]\,.
\label{nonabelian}
\eeqa
Although the situation of the triple gauge boson interaction is similar to the one 
in the Abelian case, we show the explicit 
result here because of its phenomenological relevance. For the relevant case with $(+,+)$  boundary conditions, we are interested in 
the interactions involving the zero-mode gluon with the first and second KK gluons: 
$G^{(2)}G^{(1)}G^{(0)}$. 
These can be derived from (\ref{nonabelian}) and they are of the form
\beqa
-S_{topo} &=& \frac{3\,\bar{g}_3^3}{24\pi^2\,L^{3/2}}\,\int\,\Tr\left[ G^{(2)} G^{(1)} dG^{(0)}
\right]\, \nonumber\\
&&\hspace{1cm}\times\,\left\{\left(f^{1}(L)-f^{1}(0)\right)\left(f^{2}(L)+f^{2}(0)\right) 
- 2\,\int_0^L dy f^{2}(y)\,\partial_y f^{1}(y)
\right\}\,,     
\label{threegluons}
\eeqa
where again we have used a flat profile for the zero massless mode. Computing the coefficient explicitly for the $(+,+)$ 
wave-functions results in 
\beq
-S_{topo}\simeq \frac{3\,g_3^3}{24\pi^2}\,\int\,\Tr\left[ G^{(2)} G^{(1)} dG^{(0)}\right]\, 
(-3\,k\,L)\,,
\label{threegluenumber}
\eeq  
where $g_3$ is already the 4D $SU(3)_c$ gauge coupling. There is also a quartic interaction associated with this one by gauge invariance: $G^{(2)}G^{(1)}G^{(0)}G^{(0)}$. Its coefficient is identical to the one in (\ref{threegluons}) up to a factor of $g_3$, and can be thought of as replacing the operator in (\ref{threegluons}) by the gauge-invariant combination $G^{(2)}G^{(1)}(d\,G^{(0} + g_3\,G^{(0)}\,G^{(0)})$. 
For instance, and as we will show in the next section, both these contributions must
be present when considering the process $pp\longrightarrow G^{(1)}G^{(2)}$, not only 
because they are of the same order in $g_3$ but also by requiring gauge invariance. 
Although this process involves  the second KK mode of the gluon, the fact that it is 
relatively unsuppressed makes it of phenomenological
relevance. This is specially the case for Higgsless models, where the overall 
KK-mass scale is considerably lower than in most 
other warped extra dimension scenarios. We will study the discovery potential of this 
phenomenology in the next section.

Finally, considering generic quartic interactions for $(+,+)$ KK gluons, 
there is always a trace over four $SU(3)_c$ generators given by
\beqa
\Tr[t^a\,t^b\,t^c\,t^d]= \frac{- i f_{ade}
   d_{ebc}+  i d_{ead}
   f_{bce}+  d_{ead}
   d_{ebc}- d_{ebd}
   d_{eac}+  d_{ecd}
   d_{eab}}{8} +\frac{\delta _{ad}
   \delta _{bc}  - \delta_{ac}\delta _{bd} + \delta _{ab} \delta_{cd} }{4 N_c}\,.
   \eeqa
But since the Lorentz indexes are contracted with the 
totally anti-symmetric $\epsilon$ tensor, one can at most have two identical 
KK modes in the interaction. 
As a consequence, there are no interactions like $G^{(0)}G^{(0)}G^{(0)}G^{(0)}$ 
and $G^{(0)}G^{(0)}G^{(0)}G^{(1)}$. 
%On the other hand, there will be topological interactions with total KK-number equal to 2.
\vskip0.5cm

\noindent{\bf Product Gauge Groups}

Finally, we generalize to  the case of product gauge groups. 
Specifically, we consider a $SU(3)_c\times U(1)_Y$ gauge group and two fermions with charges $(3, Y)$. 
We want to study the case when $SU(3)_c$ always has a massless mode corresponding to the 
gluon. 
We then choose $(+, +)$  boundary conditions for the $SU(3)_c$ gauge fields. 
We are particularly interested in the coupling among two $SU(3)_c$ KK modes
and one $U(1)_Y$ KK mode. The contributions from the CS terms can be read off Eq.~(\ref{csinproduct}), while the additional non-local terms are introduced following the discussion at the beginning of this section. The interaction with the lowest total KK number is 
\beqa
-S_{topo}&=&\frac{3\,Y\,\bar{g}_1\,\bar{g}_3^2}{24\,\pi^2\,L^{3/2}}\,\int \Tr\,
[B^{(0)}\,dG^{(0)}\,G^{(1)}]\,   \nonumber \\
&&\hspace{1cm}\times\,\left\{\int^L_0\,d\,y\,2\,f_G^1(y)\,\partial_5\,f_B^0(y) 
+ [f^0_B(0)\,-\,f^0_B(L)][f^1_G(0)\,+\,f^1_G(L)]
\right\}  \,,
\label{eq:B0G0G1}
\eeqa
where a constant value for $f^0_G(y)$ is used. 
Choosing $(+, +)$  boundary conditions for the  $U(1)_Y$ gauge bosons,  
we also have $f^0_B(y)$ to be $y$ independent. Then, we obtain 
\beqa
-S_{topo}&=& \frac{3\,Y\,\bar{g}_1\,\bar{g}_3^2}{24\,\pi^2\,L^{3/2}}\,\int d^4x\,
\Tr [B^{(0)}\,dG^{(0)}\,G^{(1)}]\,\times\, 0 \qquad (+, +) \,,
\label{eq:productpp}
\eeqa
reflecting the unbroken gauge symmetry. On the other hand, 
for $(+, -)$ boundary conditions for $U(1)_Y$ gauge bosons, 
the interaction is non-vanishing and is given by
\beqa
-S_{topo}&\approx& \frac{3\,Y\,\bar{g}_1\,\bar{g}_3^2}{24\,\pi^2\,L^{3/2}}\,\int d^4x\,
\Tr_3\,[B^{(0)}\,dG^{(0)}\,G^{(1)}]\,
(2\,-\,\sqrt{2})\,\sqrt{k\,L}    \nonumber \\
&=&  \frac{3\,Y\,g_1\,g_3^2}{24\,\pi^2}\,\int d^4x\,\frac{(2\,-\,\sqrt{2})}{2}\,\sqrt{k\,L}\,
\epsilon^{\mu\nu\rho\sigma}\,B^{(0)}_\mu\,
\partial_\nu\,G^{(0)}_{a, \rho}\,G^{(1)}_{a, \sigma}  \qquad (+, -) \,,
\label{eq:productpm}
\eeqa
for $k\,L  \gg 1$, where $g_i = \bar{g}_i/\sqrt{L}$  are the 4D gauge couplings. 
One can check that the coefficient of the gauge-symmetry-violating operator, 
$B^{(1)}\,dG^{(0)}\,G^{(0)}$ actually vanishes.

For quartic gauge boson interactions, we restrict ourselves to 
interactions with a total KK number below 2. This leaves only four possible interactions 
generated by the CS terms: 
$B^{(0)}G^{(0)}G^{(0)}G^{(0)}$, $B^{(1)}G^{(0)}G^{(0)}G^{(0)}$, 
$B^{(0)}G^{(0)}G^{(0)}G^{(1)}$. 
Again, we fix the boundary conditions for $SU(3)_c$  to be $(+, +)$. 
Independently of the boundary conditions for the $U(1)_Y$ field, 
the coefficients of $B^{(0)}G^{(0)}G^{(0)}G^{(0)}$ and 
$B^{(1)}G^{(0)}G^{(0)}G^{(0)}$ vanish due to the preserved gauge invariance of the 
zero-mode theory. On the other hand, the 
coefficient of $B^{(0)}\,G^{(1)}\,G^{(0)}\,G^{(0)}$ is given by the expression
\beqa
-S_{topo}&=&\frac{3\,Y\,\bar{g}_1\,\bar{g}_3^3}{24\,\pi^2\,L^{2}}\,\int d^4 x\,\Tr_3\,
[B^{(0)}\,G^{(1)}\,G^{(0)}\,G^{(0)}]\,   \nonumber \\
&&\hspace{1cm}\times\,\left\{\int^L_0\,d\,y\,2\,f_G^1(y)\,\partial_5\,f_B^0(y) 
+ [f^0_B(0)\,-\,f^0_B(L)][f^1_G(0)\,+\,f^1_G(L)]
\right\}  \,,
\label{eq:B0G1G0G0}
\eeqa
which has a coefficient  identical to the one of  $B^{(0)}dG^{(0)}G^{(1)}$ in 
Eq.~(\ref{eq:B0G0G1}). Once again, this is a consequence of the $SU(3)_c$ gauge symmetry. Adding  
Eq.~(\ref{eq:B0G0G1}) and  Eq.~(\ref{eq:B0G1G0G0}), we obtain the  $SU(3)_c$ 
gauge-invariant operator  
$B^{(0)}\,G^{(1)}\,(dG^{(0)}+g_3\,G^{(0)}G^{(0)})$. Explicitly, the quartic coupling
coefficient is, for the $(+,+)$ boundary conditions for the $U(1)_Y$ field 
\beqa
-S_{topo}&=& \frac{3\,Y\,\bar{g}_1\,\bar{g}_3^3}{24\,\pi^2\,L^{2}}\,\int d^4x\,\Tr_3\,
[B^{(0)}\,G^{(1)}\,G^{(0)}\,G^{(0)}]\,\times\, 0 \qquad (+, +) \,,
\eeqa
whereas for the $(+,-)$ boundary conditions is given by 
\beqa
-S_{topo}&\approx& \frac{3\,Y\,\bar{g}_1\,\bar{g}_3^3}{24\,\pi^2\,L^{2}}\,
\int d^4x\,\Tr_3\,[B^{(0)}\,G^{(1)}\,G^{(0)}\,G^{(0)}]\,(2\,-\,\sqrt{2})\,\sqrt{k\,L}    
\nonumber \\
&=&  \frac{3\,Y\,g_1\,g_3^3}{24\,\pi^2}\,\int d^4x\,\frac{i\,(2\,-\,\sqrt{2})}{4}\,
\sqrt{k\,L}\,\epsilon^{\mu\nu\rho\sigma}\,f^{abc}\,B^{(0)}_\mu\,G^{(1)}_{a, 
\nu}\,G^{(0)}_{b, \rho}\,G^{(0)}_{c, \sigma}  \qquad (+, -) \,,
\eeqa

We can easily  obtain the coefficient for the 
interaction $B^{(1)}\,G^{(1)}\,G^{(0)}\,G^{(0)}$  by replacing $f^0_B$ with $f^1_B$  
in Eq.~(\ref{eq:B0G1G0G0}). 
Once again, for $(+, +)$ boundary conditions for the $U(1)_Y$ the coefficient of  
$B^{(1)}\,dG^{(0)}\,G^{(1)}$ vanishes,  
whereas for $(+, -)$  boundary conditions, their coefficients are non-zero.

To summarize, in order to have topological interactions among gauge 
bosons with $(+, +)$ boundary conditions 
one needs to include the 2$^{nd}$-KK mode.
This is generically the 
case in the warped extra dimension models with gauge bosons and fermions propagating in the 
bulk. 
The interactions of phenomenological interest with the lowest KK number were derived from the master expression Eq.~(\ref{eq:master}), and are given 
by Eq.~(\ref{threegluenumber}) for the case of Non-Abelian gauge groups. All interactions involving only the zero mode and first KK mode with $(++)$ 
boundary conditions vanish, as shown in Eqs.~(\ref{eq:abelianpp}) and (\ref{eq:productpp}). 
Since the current constraints on the compactification scale are such that the 
2$^{nd}$-KK modes for gauge bosons must have masses above $\sim 5$~TeV, we do not anticipate 
that the topological interactions in this model can be discovered at the early stages of the LHC.
 
It is possible to evade the need for the 2$^{nd}$-KK mode in warped extra dimension models
where at least one of the gauge fields has $(+,-)$  boundary conditions. 
For the case of an Abelian gauge group this leads to a non-vanishing interactions involving two zero modes
and a KK mode as shown in Eq.~(\ref{eq:abelianpm}), whereas a similar expression for the case of $SU(3)\times_U(1)_Y$ is shown in Eq.~(\ref{eq:productpm}).
This boundary conditions are encountered, for instance, in Higgsless models where the electroweak symmetry is 
broken in the IR brane by boundary conditions.  
This leads to the appearance of non-vanishing topological interactions
in these scenarios. As an example in Higgsless models, there will be interactions
involving the first KK-gluon with a  $Z$ and a gluon, 
leading to potentially interesting new signals.  Other interactions involving
only electroweak KK gauge bosons or zero modes are also generated. 
In the next section we study the phenomenological implications of some of these 
novel interactions at the LHC.

%----------------------------------------------------------------------
\section{Phenomenology of Topological Interactions}
%----------------------------------------------------------------------
\label{pheno}

We are now in a position to study the phenomenological consequences of the remnant topological interactions 
discussed above. We specifically consider three types of scenarios with  
a warped extra dimension: 
Higgsless models, typical bulk warped models with a light Higgs localized near the IR brane, 
and finally the model of Ref.~\cite{Agashe:2007jb} with an implementation of KK parity, which we show is broken by the remnant  topological interactions.  
Since the form of the topological interactions depends on the details of the zero-mode localization in the 5D bulk,  
we will obtain them in the simplified picture where $t_R$ is completely IR-localized, and all other zero modes, including 
$t_L$, are localized on the UV brane. As shown in Section~\ref{wzwtocs}, in this setup only $t_L$ and $t_R$ contribute to the
topological interactions. We hope that this schematic approximation will give  a good estimate of the correct answer. 
In realistic warped extra dimension models both $t_R$ and $t_L$ are less localized, so we expect that the top quark contribution to the 
topological interactions will be somewhat smaller. On the other hand, we also neglect the $b$-quark contributions which could be comparable 
if $b_L$ is far from the UV brane where $b_R$ is assumed to be localized. 
We will address the corrections to these 
approximations in Ref.~\cite{paper2}, where we will present the most general form of the 
topological interactions.

Although in general, as it was shown in Sections~\ref{wzwtocs} and \ref{topproc}, two KK modes are needed if the gauge symmetry is unbroken. The breaking of the electroweak symmetry either by boundary conditions or by a Higgs VEV allows for effective interactions not involving the second KK mode. We also study the interactions involving a zero-mode gluon with the first two KK modes, since it is present in all models and it has the largest possible coefficient among the topological interactions. This is particularly important in Higgsless models, where the overall KK-mass scale is smaller than in other cases. 

\subsection{Higgsless Models}
Here we study the topological interactions in 
Higgsless models. 
In the Higgsless scenario~\cite{Csaki:2003zu}, the gauge symmetry in the 
bulk is $SU(2)_L\times SU(2)_R\times U(1)_{B-L}$ with $A_M^{L\,a}$, $A_M^{R\,a}$ and $B_M$ 
as their gauge bosons.  On the Planck brane, $SU(2)_R\times U(1)_{B-L}$ breaks down to 
$U(1)_Y$ hypercharge. On the TeV brane, $SU(2)_L\times SU(2)_R$ breaks to $SU(2)_D$. 
So, the final unbroken symmetry is only $U(1)_{em}$. In this model, the $W$ and $Z$ 
gauge bosons have masses  given by
\beqa
M^2_W\,=\,\frac{k^2\,e^{-2\,k\,L}}{k\,L}\,,\qquad\qquad  
M^2_Z\,=\,\frac{g_5^2\,+\,2\,\tilde{g}_5^2}{g_5^2\,+\,\tilde{g}_5^2}\,
\frac{k^2\,e^{-2\,k\,L}}{k\,L}\,.
\eeqa
Here, $g_5$ is the gauge coupling of the two $SU(2)$'s and $\tilde{g}_5$ is the gauge coupling of $U(1)_{B-L}$. 
The physical $W$ gauge boson determines one combination of parameters: $k$ and $L$. Up to leading order in $1/(kL)$, 
the relation between the 5D and the 4D gauge couplings is
\beqa
g^2\,=\,\frac{g_5^2}{L}\,,\qquad  g^{\prime 2}\,=\,\frac{g_5^2\,\tilde{g}_5^2}{({g_5^2\,+\,\tilde{g}_5^2})\,L}\,,
\qquad  e^2\,=\,\frac{g_5^2\,\tilde{g}_5^2}{({g_5^2\,+\,2\,\tilde{g}_5^2})\,L}\,.
\eeqa
The presence of the $SU(2)_R$ ensures that the $\rho$ parameter is one at leading order in $1/(kL)$.  

We first consider the topological interactions involving a $Z$, a gluon and a KK gluon. These are originated 
by CS terms containing two 5D gluon fields and a $U(1)_{B-L}$ field. These interactions are made possible in Higgsless
models since the part of the $Z$ that comes from the $U(1)_Y$ gauge boson has a non-zero mode component. 
In this model, all the three neutral 
gauge bosons: $B$, $A^{L\,3}$ and $A^{R\,3}$ contain the physical $Z$ boson. The fraction for the $B$ gauge boson is approximately
\beqa
f^{(Z)}_B(y) \simeq -\sqrt{\frac{g^2_5+\tilde{g}^2_5}{g^2_5+2\tilde{g}^2_5}}\,\frac{g_5\tilde{g}_5}{g_5^2+\tilde{g}^2_5}\,\Bigl [ 1 + 
\frac{g^2_5+2\tilde{g}^2_5}{g^2_5+\tilde{g}^2_5}\,\frac{L -y}{2L}\,e^{-2k(L-y)} \Bigr ]\,,
\eeqa
which is almost flat. This is because the $Z$ boson is mainly contained in $A^{L\,3}$ and $A^{R\,3}$, 
one linear combination of which has a ``$-$" boundary condition on the IR brane. Substituting $f^{(Z)}_B(y)$ into Eq.~(\ref{eq:B0G0G1}), 
we obtain the coefficient of the $Z G^{(0)}G^{(1)}$ contribution 
\beqa
\int^L_0\,d\,y\,2\,f_G^1(y)\,\partial_5\,f_B^{(Z)}(y) 
+ [f^{(Z)}_B(0)\,-\,f^{(Z)}_B(L)][f^1_G(0)\,+\,f^1_G(L)]
  \approx  \sqrt{\frac{g^2_5+2\tilde{g}^2_5}{g^2_5+\tilde{g}^2_5}}\,\frac{g_5\tilde{g}_5}{g_5^2+\tilde{g}^2_5}\,\frac{1}{5\sqrt{kL}}\,.
\eeqa
In this way the full  topological interaction has the form 
\beqa
-S_{topo} &=& \frac{3\,Q_{\rm sum}\,\tilde{g}\,g_3^2}{24\,\pi^2}\,\int d^4x\, 
\sqrt{\frac{g^2_5+2\tilde{g}^2_5}{g^2_5+\tilde{g}^2_5}}\,\frac{g_5\tilde{g}_5}{g_5^2+\tilde{g}^2_5}\,
\frac{1}{5\sqrt{kL}}\,\epsilon^{\mu\nu\rho\sigma}\,\frac{1}{4}\,Z_\mu\,G^{(0)}_{a, \nu \rho}\,G^{(1)}_{a, \sigma}   \nonumber \\
&=& \frac{3\,Q_{\rm sum}\,e\,g_3^2}{24\,\pi^2}\,\int d^4x\, \frac{\cos{\theta_W}}{\sin^3{\theta_W}}\,
\frac{1}{5\sqrt{kL}}\,\epsilon^{\mu\nu\rho\sigma}\,\frac{1}{4}\,Z_\mu\,G^{(0)}_{a, \nu \rho}\,G^{(1)}_{a, \sigma}  \nonumber \\
&\equiv& {\cal F}\,\int d^4x\, \epsilon^{\mu\nu\rho\sigma}\,\frac{1}{2}\,Z_\mu\,G^{(0)}_{a, \nu \rho}\,G^{(1)}_{a, \sigma} 
\,,
\label{eq:higgslesscoupling}
\eeqa
where we have used the relation between 5D and 4D couplings, and $Q_{\rm sum}$ adds all the $U(1)_{B-L}$ charge contributions and depends
on the choice of fermion representation in the 5D bulk.
In order to give an estimate in a  specific example, we choose the bulk fermions to transform as   
$Q_L = (t_L, b_L)^T \sim (3, 2,  1)_{1/6}$ and   $Q_R = (t_R, b'_R)^T \sim (3, 1, 2)_{1/6}$, resulting in $Q_{\rm sum}= (1/6+1/6) =1/3$.
The mass of $G^{(1)}$ in this model is given by $M_{G^{(1)}}\approx x_1\, k \,e^{-kL} = 2.45 \sqrt{k\,L}\,M_W$, 
which is approximately 1.2~TeV. From this action, we can read the Feynman diagram for triple gauge boson and quartic gauge boson couplings.
\begin{figure}[ht!]
\vspace{0.3cm}
\centerline{ \hspace*{0.0cm}
\includegraphics[width=0.45\textwidth]{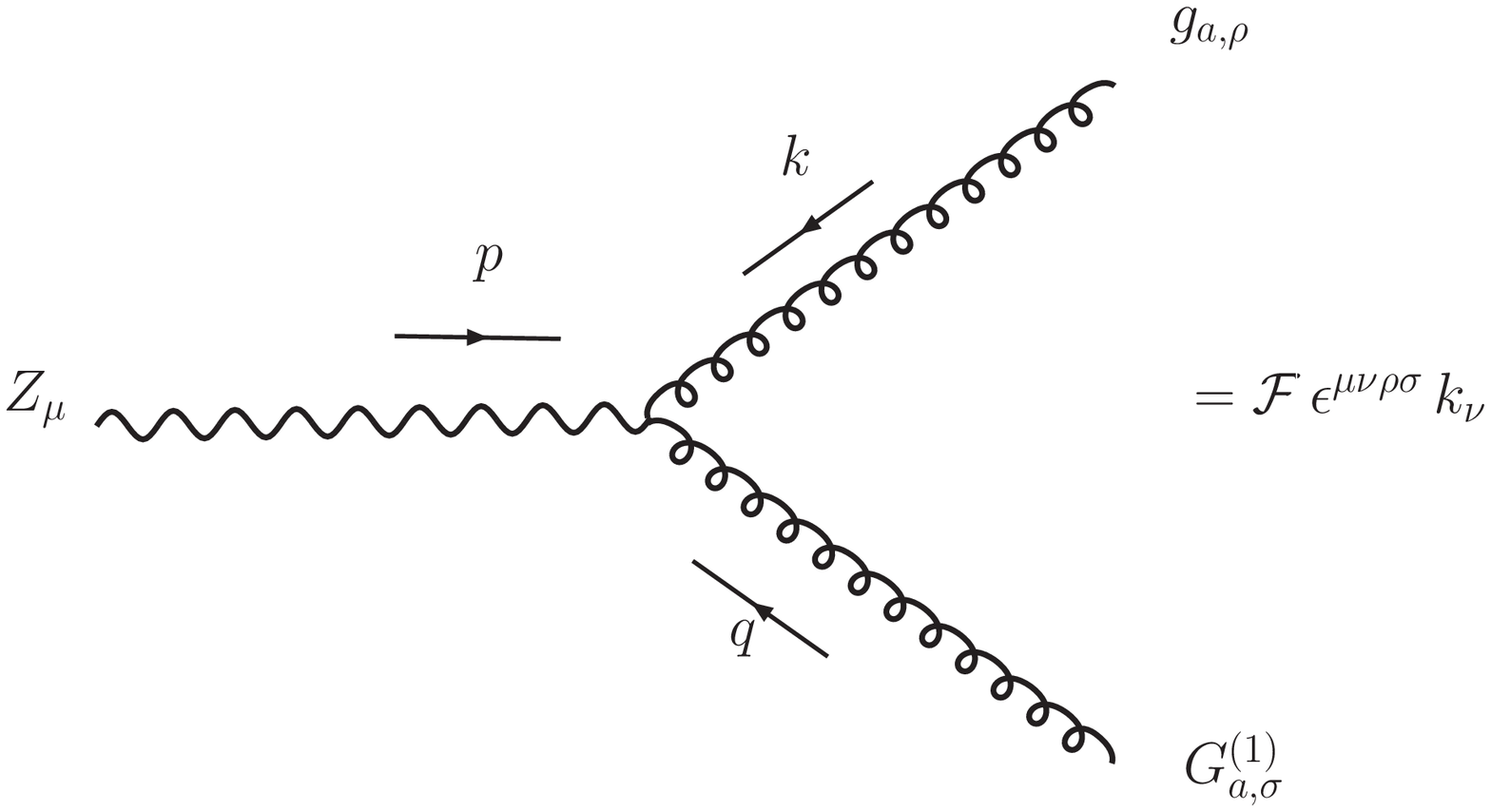} \qquad
\includegraphics[width=0.48\textwidth]{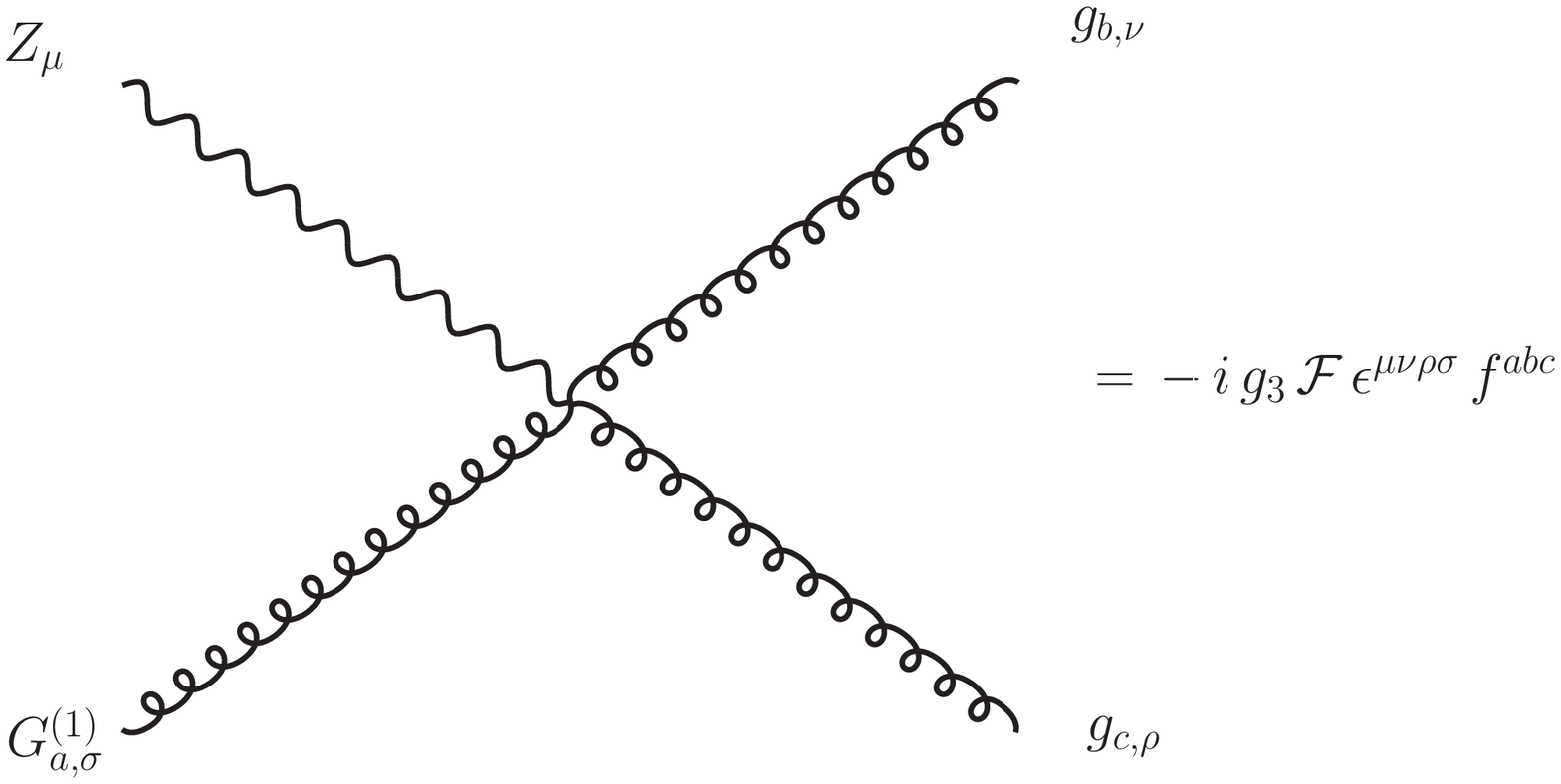}
}
\caption{\em Feynman diagrams for triple and quartic gauge boson interactions from 
topological interactions in the Higgsless model. 
The coefficient ${\cal F}$ is defined in Eq.~(\ref{eq:higgslesscoupling}).
}
\label{fig:higgslessFeyman}
\end{figure}
To estimate the size of the effect we look first at the partial width for the 
newly induced decay channel for the first KK-gluon via the topological interaction. 
Using the triple gauge boson coupling in Fig.~\ref{fig:higgslessFeyman}, the topological decay width of $G^{(1)}$ is given
\beqa
\Gamma(G^{(1)}\rightarrow Z + g) \,=\,\frac{1}{96\,\pi}\,\frac{M^3_{G^{(1)}}}{M_Z^2}\,{\cal F}^2 \,=\,M_{G^{(1)}}\,
\frac{x_1^2\,Q^2_{\rm sum}\,\alpha\,\alpha_c^2}{9600\,\pi^2}\,\frac{\cos^4{\theta_W}}{\sin^6{\theta_W}}\,\approx\,10^{-7}\,M_{G^{(1)}}\,.
\eeqa
where the longitudinal enhancement of the $Z$ somewhat compensates the $1/\sqrt{k L}$ 
suppression factor in the coupling. However, this partial width is still extremely small, implying that  the topological decay 
channel cannot compete with the fermionic decay modes, 
unless all fermions have  a flat profile and therefore have highly suppressed couplings 
to the first KK-gluon. As a result, processes involving this interaction at the LHC such as 
$pp\to G^{(1)} Z$, have a very suppressed production cross section.

\begin{figure}[ht!]
\vspace{0.3cm}
\centerline{ \hspace*{0.0cm}
\includegraphics[width=0.45\textwidth]{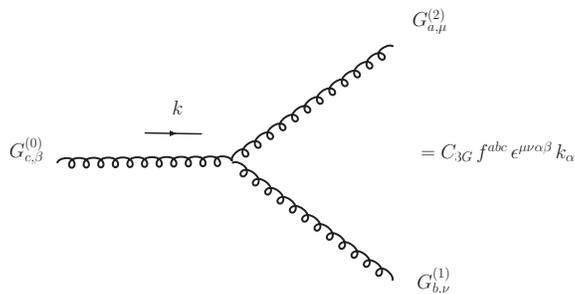} \qquad
}
\caption{\em Feynman diagram for the triple interaction $G^{(0)}G^{(1)}G^{(2)}$.
The coefficient $C_{3G}$ is defined in (\ref{threeg}).
}
\label{threegluediagram}
\end{figure}

\begin{figure}[ht!]
\vspace{0.3cm}
\centerline{ \hspace*{0.0cm}
\includegraphics[width=0.6\textwidth]{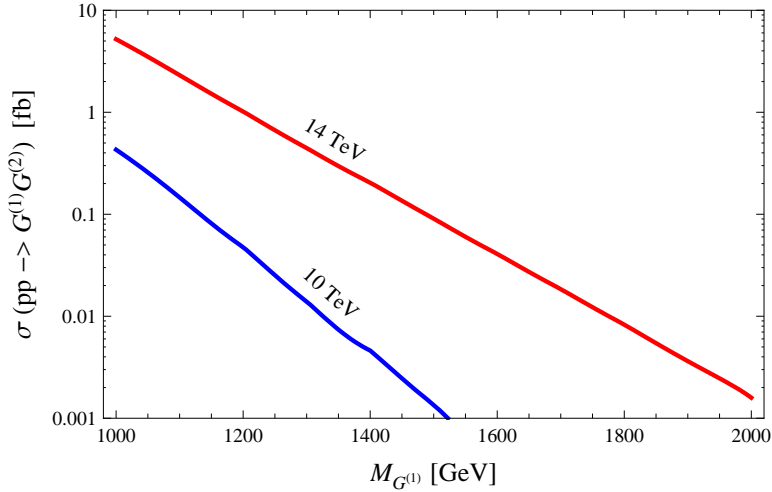} 
}
\caption{\em The production cross section of $pp\to G^{(1)}+G^{(2)}$ as a 
function of the first KK gluon mass at the LHC, for $\sqrt{s}=14$ and $10~$TeV. 
Only the $q\bar q$ contributions are 
taken into account. This result applies to all warped extra dimension models as long 
as $SU(3)_c$ is a bulk gauge symmetry and is independent of the details of the fermion 
sector. 
}
\label{fig:productionCS}
\end{figure}

Also of interest is to compute the effects driven by the $\left[SU(3)_c\right]^3$ CS term 
with the lowest possible KK number. This corresponds to the 
$G^{(0)}G^{(1)}G^{(2)}$ interaction as 
discussed in Section~\ref{topproc}. As discussed there, this interaction is the least suppressed one and is 
present in all  models. It is more relevant 
in Higgsless models since in them the KK-mass scale needs to be  lower 
than in more generic warped extra dimension 
theories. Therefore, the presence of the 
second KK gluon in the interaction 
may not necessarily preclude its observation at the LHC.   
From the triple interaction in Eq.~(\ref{threegluenumber}) we obtain the Feynman rule depicted in Figure~\ref{threegluediagram}, with the 
coefficient defined as 
\beq
C_{3G} = \frac{g_3^3}{8\pi^2}\, N_f\,\frac{3}{4}\,kL~,
\label{threeg}
\eeq
where $N_f$ is the number of chiral colored fermions contributing to the appropriate CS term. 
In our case, $N_f=2$, since there are two left-handed anomalies associated 
with the $Q_R$ multiplet. The coupling is large enough that it results in a non-negligible 
contribution to the $G^{(2)}$ decay width. This is given by
\bea
\Gamma(G^{(2)}\to G^{(1)} + g)&=&\frac{C^2_{3G}}{32\,\pi}\, 
\frac{\left(M^2_{G^{(2)}}\,+\,M^2_{G^{(1)}}\right)\left(M^2_{G^{(2)}}\,-\,M^2_{G^{(1)}}\right)^3}{M^2_{G^{(1)}}\,M^5_{G^{(2)}}}\,,  \nonumber \\
&=& \frac{9\,\alpha_c^3\,N_f^2\,k^2\,L^2}{512\,\pi}\,\frac{\left(x_2^2\,+\,x_1^2\right)\left(x^2_2\,-\,x_1^2\right)^3}{x_1^2\,x_2^2}\,M_{G^{(2)}}\,,
\label{g2width}
\eea
with $x_1\approx2.45$ and $x_2\approx 5.56$, obtained from the roots of Bessel functions. 
For instance, for $M_{G^{(1)}}=1.2~$TeV, corresponding to $M_{G^{(2)}}=2.7~$TeV, 
evaluating $\alpha_s(M_{G^{(2)}})$, for $N_f=2$ and $kL\approx 37.5$, we have 
\beq
\Gamma(G^{(2)}\to G^{(1)} + g)\,\approx\,0.02 \,M_{G^{(2)}}\,.
\eeq
Thus, we see that this decay mode of the second KK mode of the gluon induced by 
topological interactions is 
significant, due to the enhancement from the 
large wave-function overlapping factor $k\,L\sim 35$.
We can compare this decay channel of $G^{(2)}$ with its decays to fermions. Although
these are more  model dependent, we can estimate  them 
by making use of the dominant fermion decay channel, $G^{(2)}\to t_R\bar t_R$, since 
$t_R$ is the most IR-localized zero-mode quark. Assuming a value for the bulk mass parameter $c_{t_R}$ large enough to be consistent with our approximate calculations for the topological interactions, the branching fraction 
into $G^{(1)} g$ is of the order of a few percent. For instance, for $c_{t_R}\simeq 3$ we have 
$Br(G^{(2)}\to G^{(1)} + g)\simeq 0.04$. 
%In Figure~\ref{fig:g2decays} we plot the relative branching ratios for the 
%$G^{(2)}$ decays into $t_R\bar t_R$ and $G^{(1)}$. The decay through topological 
%interactions dominates for a wide range of values of $t_R$ localization consistent
%with obtaining the correct value for $m_t$. Even for values of the $t_R$ localization 
%parameter $c_R$ for which the fermion decay dominates, the $G^{(1)} + g$ channel
%remains almost as important and therefore should be also observable. 
%
%\begin{figure}[ht!]
%\vspace{0.3cm}
%\centerline{ \hspace*{0.0cm}
%\includegraphics[width=0.6\textwidth]{g2brs.eps} 
%}
%\caption{\em Branching ratios 
%for $G^{(2)}$ decays as a function of the $t_R$ localization parameter $c_R$,
%assuming $G^{(1)} g$ and $t_R\bar t_R$ as the only two decay channels.
%}
%\label{fig:g2decays}
%\end{figure}
%
The coupling $G^{(0)}G^{(1)}G^{(2)}$ is also large enough so as to make it interesting to 
estimate the cross section for $pp\to G^{(1)}G^{(2)}$ 
induced by this interaction. Given that the invariant mass of these events are larger 
than $1~$TeV, it is enough 
to use $q\bar q\to G^{(1)}G^{(2)}$ to estimate this cross section, since the quark 
parton distribution functions 
are dominant in this energy regime. We obtain 
\bea
\sigma(q\bar q\to G^{(1)}G^{(2)})&=&\frac{C^2_{3G}\,g_3^2}{72\,\pi\,\hat{s}^{3}\,
M^2_{G^{(1)}}\,M^2_{G^{(2)}}}\,\sqrt{(\hat{s}\,
+\,M^2_{G^{(1)}}\,-\,M^2_{G^{(2)}})^2\,-\,4\,\hat{s}\,M^2_{G^{(1)}}}    \\
&& \hspace{-3.5cm} \times \left[   M^2_{G^{(1)}} (8\,\hat{s}\,M^2_{G^{(2)}}\,-\,M^4_{G^{(2)}}\,+\,\hat{s}^2)\,-
\,M^4_{G^{(1)}}(M^2_{G^{(2)}}\,+\,2\,\hat{s})\,+\,M^6_{G^{(1)}}\,+\,(M^3_{G^{(2)}}\,-\,\hat{s}\,M_{G^{(2)}})^2      \right]\,,   \nonumber 
\eea
which, for $M_{G^{(1)}}=1.2~$TeV results in $\sigma(pp\to G^{(1)}G^{(2)})
\simeq 1~{\rm fb}$. We show the production cross section at LHC 
for two different center of mass energies in Fig.~\ref{fig:productionCS}, as a 
function of the mass of the first KK gluon.
We can see that for $M_{G^{(1)}}$ masses consistent with Higgsless models, 
a production cross section of several fb can be obtained at the LHC with 
$\sqrt{s}=14~$TeV. For the heavier masses typically required in 
warped extra dimension models with a light Higgs, the cross section
drops considerably below $0.1~$fb, making its observation at the LHC 
very challenging. 
We emphasize that these interactions are present in all warped extra 
dimension models, and their strength is very model-independent. 
However, the presence of the second KK gluon makes them only 
relevant in Higgsless models, since the higher KK-mass scale in other scenarios 
makes its production cross section too small for early observation at the LHC. 

%----------------------------------------------------------------------
\subsection{The Standard Warped Extra Dimension Model}
%----------------------------------------------------------------------
In this section, we address the traditional warped extra dimension scenarios, which typically have an IR-localized light Higgs. 
The gauge symmetry in the bulk is $SU(2)_L\times SU(2)_R\times U(1)_X$, with the 
gauge bosons having 
$(+, +)$ boundary conditions. 
Choosing the same gauge coupling for $SU(2)_L$ and $SU(2)_R$  
as $g_L$ and the gauge coupling for $U(1)_X$ as $g_X$, we have
\beqa
g^\prime = \frac{g_X\,g_L}{\sqrt{g_L^2\,+\,g_X^2}}\,,\qquad\, e = \frac{g_L\,g^\prime}{\sqrt{g^{\prime 2}\,+\,g_L^2}} \,, 
\quad {\rm or}\quad g_X=\frac{e\,g^\prime}{\sqrt{2\,e^2\,-\,g^{\prime 2}}}\,,\quad g_L=\frac{e\,g^\prime}{\sqrt{g^{\prime 2}\,-\,e^2}}\,.
\eeqa

The first topological interaction involving two KK modes is  $X^{(2)}G^{(0)}G^{(1)}$. 
Substituting $f^{(2)}_X(y)$ into Eq.~(\ref{eq:B0G0G1}), we have the wave-function overlapping part as
\beqa
\int^L_0\,d\,y\,2\,f_G^1(y)\,\partial_5\,f_X^{2}(y) 
+ [f^{2}_X(0)\,-\,f^{2}_X(L)][f^1_G(0)\,+\,f^1_G(L)]
 \, \approx \, -\,3.0\,k\,L\,.
\eeqa
After electroweak symmetry breaking, the physical $Z$ boson mainly contains zero modes with a small fraction in the higher KK-modes. 
The mixing angle between the $Z$ boson and $X^{(2)}$ can be written as~\cite{Agashe:2007ki}
\beq
\sin{\theta_{02X}}\,\approx\,-\frac{M_Z^2}{M^2_{X^{(2)}}}\,\sqrt{2\,\cos^2{\theta_W}-1}\,\sqrt{k\,L}\,.
\eeq
Here $M_{X^{(2)}}\approx 5.57\,k\,e^{-k\,L}\approx 2.3\,M_{G^{(1)}}$. For $M_{G^{(1)}}=2$~TeV and $kL=34$, we have $\sin{\theta_{02X}} 
\approx 0.0024$. Similar to the Higgsless model, we have ${\cal F}_{RS}$ given by
\beqa
{\cal F}_{RS}\,=\,\frac{3\,Q_{\rm sum}\,g_X\,g_3^2}{24\,\pi^2}\,\frac{(-3.0)\,k\,L}{2}\,\sin{\theta_{02X}}\,
=\,\frac{3\,Q_{\rm sum}\,e\,g_3^2}{24\,\pi^2}\,\frac{3.0\,k\,L\,\sqrt{k\,L}}{2}\,\frac{M_Z^2}{2.3^2\,M^2_{G^{(1)}}}\,.
\eeqa
The topological decay width of $G^{(1)}$ in this model is 
\beqa
\Gamma(G^{(1)}\rightarrow Z + g) \,=\,\frac{1}{96\,\pi}\,\frac{M^3_{G^{(1)}}}{M_Z^2}\,{\cal F}_{RS}^2 \,=\,M_{G^{(1)}}\,
\frac{0.08\,Q^2_{\rm sum}\,\alpha\,\alpha_c^2\,(kL)^3\,M_Z^2}{96\,\pi^2\,M^2_{G^{(1)}}}\,\approx\,2\times 10^{-7}\,M_{G^{(1)}}\,,
\eeqa
for $Q_{\rm sum} =1$, $kL =34$ and $M_{G^{(1)}}=2$~TeV.
Thus, we conclude that in warped extra dimension scenarios with an IR-localized 
Higgs these topological interactions are 
very suppressed, implying that their observation at colliders would require 
luminosities larger than the ones to be achieved at the 
LHC. Although the interaction $g G^{(1)}G^{(2)}$ is also present in this model, unlike in Higgsless 
models in the previous section its effects are highly suppressed by the fact that the $G^{(2)}$ mass exceeds $5~$TeV, pushing the associated production 
of $G^{(1)}$ and $G^{(2)}$ out of the reach of the LHC even though the couplings 
are unsuppressed. 

%----------------------------------------------------------------------
\subsection{The Warped Extra Dimension Model with KK Parity}
\label{WEDwithKKparity}
%----------------------------------------------------------------------
As a last  phenomenological application, we consider the consequences of topological interactions in a warped extra dimension model with KK 
parity, as proposed in Ref.~\cite{Agashe:2007jb}. In these model, KK parity conservation results in the 
stability of the lightest KK-odd particle making it an interesting candidate for dark matter. 
We examine the effects of the topological interactions on KK parity, and therefore on the stability of the lightest KK-odd particle.  
We will show that KK parity survives the presence of the topological interactions obtained by integrating 
heavy fermions. Since the way this comes about is non-trivial, and is in contrast to what happens in the Little 
Higgs models with T parity~\cite{Hill:2007zv}, where the symmetry is broken by the topological interactions, 
the proof warrants a fair amount of detail.    
Here we study the gravitationally stable model of Ref.~\cite{Agashe:2007jb}, which puts the UV brane at the fixed point of 
a $Z_2$ reflection in the compact dimension. 
Thus, with the extra dimension defined in the interval $y\,\in\,\left[-L, L\right]$,  the UV brane is at the origin, 
and there are two IR branes at $-L$ and $L$. The warp factor is symmetric under a $Z_2$ reflection. 
The doubling of the physical space implies the existence of 
twice the KK modes, which are now even or odd under the $Z_2$ reflection symmetry. If the $Z_2$ symmetry   
is preserved, so is KK parity. 

\begin{figure}[ht!]
\vspace{0.3cm}
\centerline{ \hspace*{0.0cm}
\includegraphics[width=0.75\textwidth]{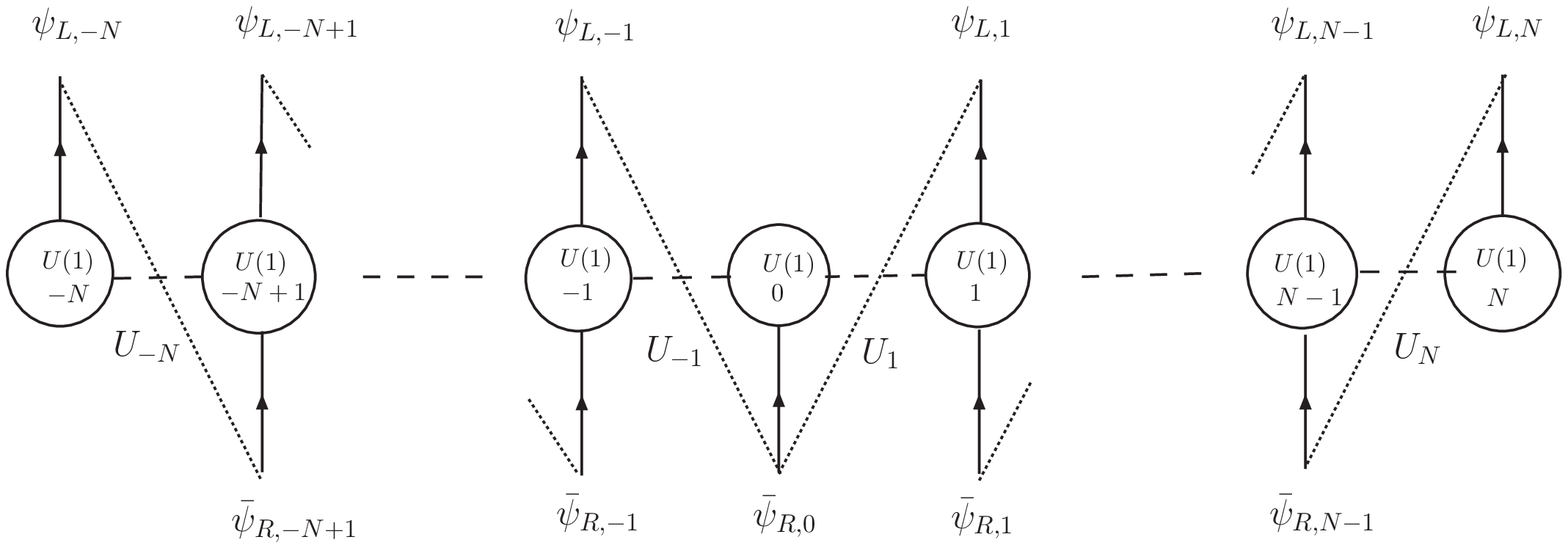}
}
\vspace{1cm}
\centerline{ \hspace*{0.0cm}
\includegraphics[width=0.75\textwidth]{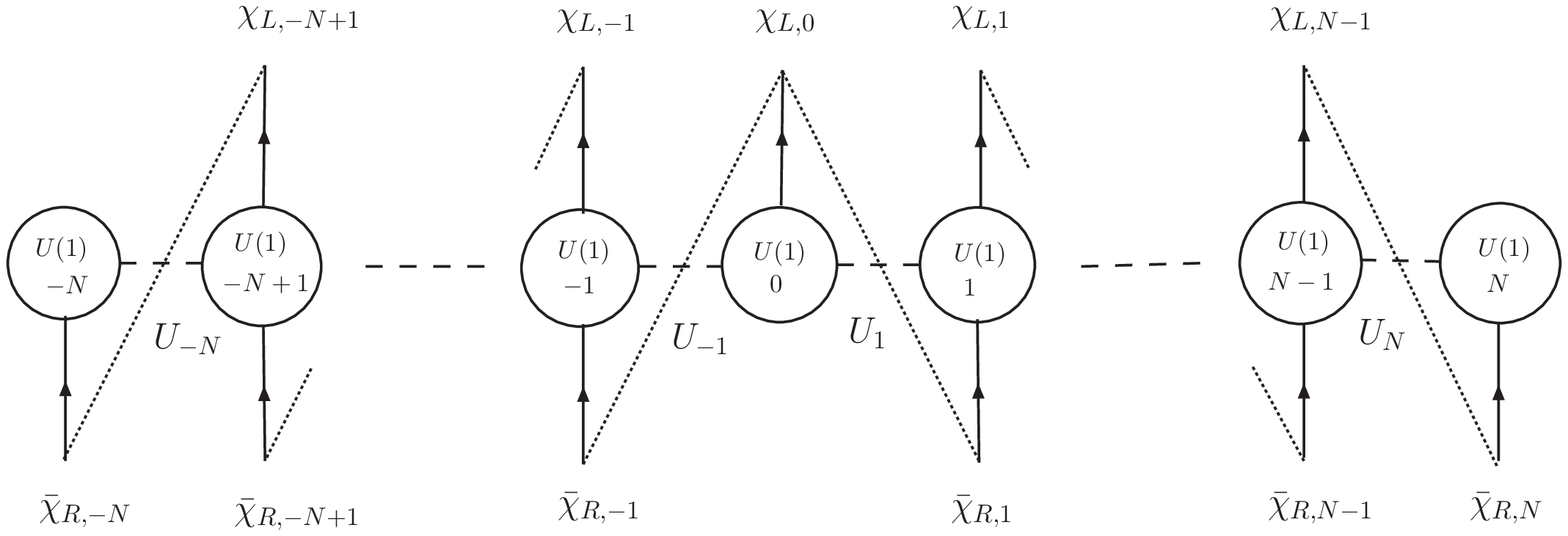}
}
\caption{\em Deconstruction of the warped extra dimension with KK parity. 
The moose diagram in the upper panel results in a left-handed zero mode, where from left to right the site number goes  
from $-N$ to $N$. We remove the left-handed mode in the zeroth site and two right-handed modes in the $-N$-th and $N$-th sites. 
A similar diagram is shown in the lower panel to obtain a right-handed zero mode.  The anomalies are canceled in each site.
The deconstruction is manifestly $Z_2$-symmetric. 
\label{fig:mooseKKparity}
}
\end{figure}

In order to study the presence of topological interactions in these models, we first deconstruct this IR-UV-IR model. 
For the purpose of discussing the topological interactions, we will not consider brane-localized terms, which are only 
introduced in order to separate the masses of KK-even and KK-odd gauge bosons. This simplification does not affect our results.
The moose diagrams corresponding to  
this model are shown in Fig.~\ref{fig:mooseKKparity}, where the link field is $U_j$ and the fermion site masses are $\mu_j$ 
with $j = -N , -N +1, \cdots , 0 , \cdots\, N-1, N$.
For fermions providing a massless left-handed zero mode (the upper panel of Fig.~\ref{fig:mooseKKparity}), 
KK-parity is defined in the continuum theory 
as $y\rightarrow -y$ with $\psi_{L, R}\rightarrow \gamma_5\,\psi_{L, R}$. 
In the deconstructed theory, it becomes $\psi_{j, L}\rightarrow -\psi_{-j, L}$ and $\psi_{j, R}\rightarrow \psi_{-j, R}$. 
To preserve KK-parity, we then require the following conditions: $U_{-j} = - U_{j}$ and  $\mu_{-j} = - \mu_{j}$, 
with $\langle U_j \rangle = v\,q^j/\sqrt{2}$ and $\mu_j = - g\,v\,q^{c_L+j-1/2}$ as in Eq.~(\ref{eq:conditions}). 
Defining $\psi^{\pm}_{L, j} \equiv (\psi_{L, j} \pm \psi_{L, -j})/\sqrt{2}$, $\psi^{\pm}_{R, j}\equiv (\psi_{R, j}\mp\psi_{R, -j})/\sqrt{2}$ 
and $\psi^{-}_{R, 0} = \psi_{R, 0}$, we have $N$ KK-even left-handed modes $\psi^+_{L, j = 1, 2, \cdots , N}$, $(N-1)$ KK-even right-handed modes 
$\psi^+_{R, j=1, 2\, \cdots , N-1}$, $N$ KK-odd 
left-handed modes $\psi^-_{L, j = 1, 2, \cdots , N}$ 
and $N$ KK-odd right-handed modes $\psi^-_{R, j = 0, 1,  \cdots , N-1}$. 
Therefore, there is one massless KK-even left-handed zero mode. 
Following the discussions of Section~\ref{sec:fermiondeconstruct}, it can be checked  
that the equations of motion, 
spectra and wave-functions of fermions indeed match the results in the continuum theory of Ref.~\cite{Agashe:2007jb}. 
For the lower panel, which provides a massless right-handed zero mode, 
one can also match the continuum results with $\mu_j = - g\,v\,q^{-c_R+j-1/2}$ and $\chi^{\pm}_{L, j} \equiv (\chi_{L, j} \mp \chi_{L, -j})/\sqrt{2}$ 
and $\chi^{\pm}_{R, j}\equiv (\chi_{R, j}\pm \chi_{R, -j})/\sqrt{2}$. 
In Table.~\ref{tab:extremlimitKKparity} we show the different limits leading to extreme localizations of the zero modes  on one of the branes.
\begin{table}[htdp]
\renewcommand{\arraystretch}{1.8}
\begin{center}
\begin{tabular}{ccc}
\hline
left-handed fermion & \hspace{1cm}   &right-handed fermion  \\
$c_L \gg \frac{1}{2} \; (\mbox{UV}) \quad \leftrightarrow \quad \frac{\mu_j}{v_j} \rightarrow 0$ 
&  \hspace{1cm} &  $c_R \gg - \frac{1}{2} \; (\mbox{IR}) \quad \leftrightarrow \quad \frac{\mu_j}{v_j} \rightarrow \infty$ \\ 
$c_L \ll \frac{1}{2} \; (\mbox{IR}) \quad \leftrightarrow \quad \frac{\mu_j}{v_j} \rightarrow \infty$ 
&  \hspace{1cm}  & $c_R \ll - \frac{1}{2} \; (\mbox{UV}) \quad \leftrightarrow \quad \frac{\mu_j}{v_j} \rightarrow 0$ \\ \hline 
\end{tabular}
\end{center}
\caption{\em Matching of the continuum warped extra dimension model with KK parity and the discretized theory for different limits.}
\label{tab:extremlimitKKparity}
\end{table}%

\begin{figure}[ht!]
\vspace{0.3cm}
\centerline{ \hspace*{0.0cm}
\includegraphics[width=0.5\textwidth]{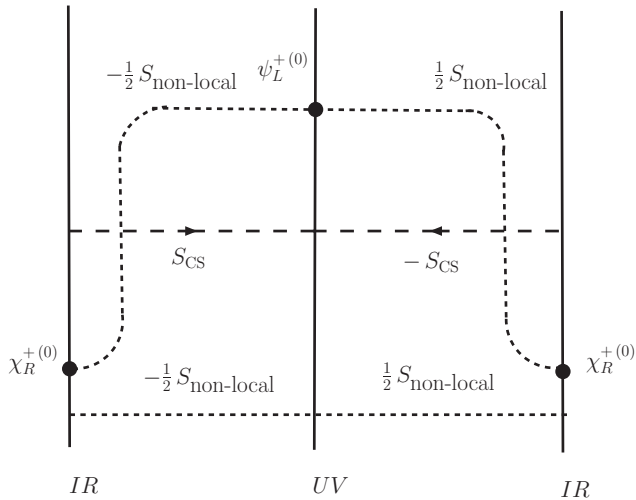}
}
\caption{\em The topological interactions in the warped extra dimension model with KK parity. 
The nonzero topological interaction is even  under KK parity.}
\label{fig:topointeractionKKparity}
\end{figure}

We want to consider the case with the two chiral zero modes localized at different extremes of the extra dimension in the continuum theory. 
In particular, just as we did earlier in this Section, we consider the situation with an IR-localized right-handed zero mode 
and a UV-localized left-handed zero mode. 
As shown in Table~\ref{tab:extremlimitKKparity} and previously discussed in Section~\ref{sec:fermiondeconstruct}, 
in order to obtain such situation we take 
the limits $\mu_j \ll v_j$ for the deconstruction with a
left-handed zero-mode, and $\mu_j \gg v_j$ for the one with a right-handed zero-mode. 
Integrating out the heavy fermions, results in a summation of WZW terms coming from the $\psi$ KK-modes.
We are also left with two left-handed modes around the zeroth site: $\psi^{+}_{L, 0}$ and $\chi^-_{L, 0}$; and 
two right-handed modes: $\chi_{R, -N}$ 
on the $-N$-th site and $\chi_{R, N}$ on the $N$'s site.  
The KK-parity odd combination $(\chi_{R, N}-\chi_{R, -N})/\sqrt{2}$ gets a Dirac mass with 
 $\chi^-_{L, 0}$. In the end, after integrating out all heavy fermions  
we have one massless left-handed mode localized on the zeroth site and one 
massless right-handed mode distributed equally on the $-N$-th site and the $N$-th site, in 
addition to the WZW terms. 
Then, when going to 
the continuum limit 
we have a theory with massless chiral fermions and KK gauge bosons, with topological 
interactions among them. These include the local CS terms resulting from 
integrating out the $\psi$ KK tower, as well as the non-local terms induced 
by the triangle diagrams with the zero modes and the odd KK modes. 
This is illustrated 
in Fig.~\ref{fig:topointeractionKKparity}: the upper non-local interactions 
are from triangular contributions from the massless fermion zero modes. 
The lower non-local interactions come from integrating out the   
$\chi^-_{L, 0}$ and $(\chi_{R, N}-\chi_{R, -N})/\sqrt{2}$. 
Summing all topological interactions and in the unitary gauge, 
we arrive at the following gauge invariant action
\beqa
-S_{topo}&=&\frac{Q^3}{24\,\pi^2}\,\int\,\epsilon(y)\,A\,d\,A\,d\,A  +\frac{Q^3}{24\,\pi^2}\,\int\,d^4x \left[A(0)\,d\,A(0)\,A(L) 
\,-\,A(L)\,d\,A(L)\,A(0)
\right]  \nonumber \\
&& \hspace{2cm} +\, \frac{Q^3}{24\,\pi^2}\,\int\,d^4x \left[A(0)\,d\,A(0)\,A(-L) \,-\,A(-L)\,d\,A(-L)\,A(0)  \right]\,,
\label{eq:topointeractionKKparity}
 \eeqa
with the function $\epsilon(y)=1$ for $y>0$ and $-1$ for $y<0$. 
Thus we see that the resulting non-vanishing  topological interactions are even 
under the $Z_2$ transformation $y \rightarrow -y$. 
Therefore, KK parity is still a good symmetry in this model. 
As a consequence, the potential dark matter candidate proposed  
in Ref.~\cite{Agashe:2007jb} would remain stable if the only topological interactions present are those
generated by integrating out the fermion spectrum as shown above. 
%{\bf However, it is not clear that KK-parity odd 
%CS terms such as 
%\begin{equation}
%\int\,A\,d\,A\,d\,A ~,
%\end{equation}
%which would be inevitably present if the compactification was equivalent to a circle, 
%are absent in this construction. Such terms would lead to the decay of the lightest KK-odd particle. }

Although the $Z_2$ symmetry in this model can remain unbroken, its origin 
does not appear to be associated with the symmetry of an orbifold geometry as it is in Universal Extra Dimensions (UED)~\cite{ued}.
In the UED case the $Z_2$ symmetry is purely geometric, and is defined about the middle point of the orbifold which 
does not play a dynamical role. Here, the symmetry is obtained by attaching two identical orbifolds at the UV brane. 
This amounts to making an assumption about the dynamics of the middle point where the UV brane is located,  
and appears to be similar to imposing an ad hoc matter parity as is done in other models.

%---------------------------------------------------------------------
\section{Conclusions}
\label{conclusions}
%----------------------------------------------------------------------
Extra dimensional theories with chiral zero modes are rendered non-anomalous by 
the addition of bulk Chern-Simons terms. Using deconstruction methods we have 
shown that these terms not only cancel the localized anomalies but also lead to 
remnant interactions among gauge KK and zero modes. In a purely four-dimensional interpretation, 
they involve the vector mesons of a global symmetry which is partially gauged. 
We derived our results in the 
limits of extreme fermion localization, which in the deconstructed language corresponds to 
taking the ratio of fermion masses to link VEVs, $\mu_j/v_j$, either to $0$ or to infinity. 
These simplifying assumptions allowed us to obtain the remnant topological interactions 
in a closed form in the continuum limit. However, it is clear from our derivation that their presence 
is a generic feature of these theories. We can also conclude from this simplified treatment that 
the topological interactions will depend on the zero-mode fermion bulk profile, as attested by the fact that 
when both chiralities are localized on the same brane there are none, whereas if the chiral zero-modes are 
localized at opposite ends of the orbifold they are present.  
The more generic case, for finite values of $\mu_j/v_j$ corresponding to 
zero-mode fermions with bulk profiles, does not lend itself to a simple form in the continuum limit, and it will be
presented elsewhere~\cite{paper2}. Here, we used our simplified result to approximate the most important 
contribution to these terms in warped extra dimension models, which comes from the zero-mode top quark. 
We consider the effects as coming from an IR-localized $t_R^{(0)}$, and a UV-localized  $t_L^{(0)}$, assuming that all other 
zero modes are UV-localized. This approximation should give a reasonable estimate of the effects in more realistic 
models of zero-mode localization. Even then, we must notice that the answer still is dependent of details of the fermion 
content of the model, such as the embedding of fermions in the 5D bulk. Such was the case  when computing the 
strength of the interactions in Section~\ref{pheno} leading to $G^{(2)}G^{(1)}g$ and $G^{(1)} Z g$ processes. 
A different embedding for the right-handed multiplets in the bulk would have led to different values of 
$Q_{\rm sum}$ and $N_f$  in (\ref{eq:higgslesscoupling})
and (\ref{threeg}) respectively. 

In the deconstruction description we showed  how for two-site models there are no 
remnant interactions, whereas already in the three-site models these are present. 
In the continuum limit, this manifest itself  in the fact that for the remnant 
interactions to be non-zero the minimum interaction must involve a zero mode, plus 
first and second KK modes, as shown in Section~\ref{topproc}. 

The remnant topological interactions lead to novel physical processes. We have shown how to derive these 
for constructions involving Abelian, Non-Abelian as well as product gauge groups relevant in various 
model-building scenarios in warped extra dimensions. 
In particular, we considered the interaction involving 
the first and second KK modes of the gluon with the gluon zero mode, $G^{(2)}G^{(1)}g$, deriving
from the CS terms that cancel the $\left[SU(3)_c\right]^3$ anomaly. This is the most un-suppressed topological interaction
in warped extra dimension models, due both to the largest possible product of gauge couplings as well as to 
the enhancement of the wave-function overlap among KK modes. 
The strength of this interaction is large enough to make it a visible decay mode of the $G^{(2)}$ for the choice of parameters used here, corresponding to a UV-localized $t_L^{(0)}$ and an IR-localized $t_R^{(0)}$. 
We also used this interaction to estimate the cross section for $pp \to G^{(1)}G^{(2)}$, plotted in Figure~\ref{fig:productionCS} 
as a function of $M_{G^{(1)}}$. This interaction is present  in all warped extra dimension models and does not depend on the details of electroweak symmetry breaking. 
On the other hand, as we can see from Figure~\ref{fig:productionCS}, the cross section is large enough to be observed at the LHC only 
for models with $M_{G^{(1)}}$ not far above $1~$TeV, as it is the case in Higgsless models.
This implies that in Higgsless models the process $pp \to G^{(1)}G^{(2)}$ induced by the topological interaction $G^{(2)}G^{(1)}g$
can be observed at the LHC. The final state would consist of four top quarks and a hard gluon jet, with 
two of the top quarks and the jet reconstructing 
to $M_{G^{(2)}}$ and the other two top quarks reconstructing to $M_{G^{(1)}}$. 

Many other processes are induced by the topological interactions. As examples, we considered the topological interactions 
induced by the $\left[SU(3)_c\right]^2\,U(1)_Y$ CS terms leading to the $ g Z G^{(1)}$ vertex. We study the phenomenology of such a 
unique interaction for different choices of boundary conditions relevant for warped models with or without a Higgs in Section~\ref{pheno}. 
We conclude that these coupling are too small to be observable at the LHC, unless the couplings of $G^{(1)}$ 
to zero-mode fermions are highly suppressed. 

Finally, we have also studied the warped extra dimension model with KK parity, as proposed in 
Ref.~\cite{Agashe:2007jb}. We showed that, unlike in the case of Little Higgs theories with T parity, 
the remnant topological interactions generated by integrating out KK fermions do not break KK parity. 
Thus, and as long as it is assumed that the 5D theory has no Chern-Simons terms, the lightest KK-odd particle 
remains stable.

Other processes can be easily derived by following the procedure presented in Section~\ref{topproc}. 
For instance, the $\left[SU(3)_c\right]^2\,U(1)_Y$ CS terms also induce the vertex $G^{(1)} Z^{(1)} g$. 
Although its coupling is smaller than that of $G^{(2)}G^{(1)} g$, the fact that it does not involve a second
KK mode may result in a phenomenologically relevant mechanism for $pp\to G^{(1)}Z^{(1)}$ production.   
Also potentially interesting are purely electroweak interactions coming from $\left[SU(2)\right]^2U(1)$ CS terms, 
resulting in couplings such as $Z^{(1)}Z^{(0)}Z^{(0)}$, which in Higgsless models are not suppressed
by wave-function factors. 

Among the processes we have not considered are those involving gravitons and their KK modes, generated 
by the cancellation of gravitational anomalies. To obtain these would require the deconstruction 
of gravity~\cite{nhm} in a warped extra dimension theory, an interesting problem in and on itself 
regardless of its phenomenological applications.
Finally, we have not made an exhaustive study of the topological interactions in 
all warped extra dimension scenarios. For instance, it would be interesting 
to consider the form of these interactions in Gauge-Higgs unification models~\cite{ghu}.

In sum, the observation of topological interactions as the ones studied here would 
point to fundamental aspects of
the physics underlying the discoveries of new massive gauge bosons at the LHC. 
We hope that this work can be the basis
for more detailed phenomenological studies of the collider signals of these topological terms.

%\end{document}

%\newpage
%%%%%%%%%%%%%%%%%%%%%%%%%%%%%%%%%%%%%%%%%%%%%%%%%%%%%%%%%%%%%%%%
\bigskip

{\bf Acknowledgments:} 
The authors are grateful to Bill Bardeen, Hsin-Chia Cheng, John Terning and 
Zhenyu Han for useful discussions and suggestions.  
G.B. acknowledges support from the John Simon Guggenheim Memorial Foundation and the 
Conselho Nacional de Desenvolvimento Cientifico e Tecnologico (CNPq). 
Y.B. and G.B thank the Aspen Center for Physics for its hospitality. 
Fermilab is operated by Fermi 
Research Alliance, LLC under contract no. DE-AC02-07CH11359 with the United States Department of Energy.  

%%%%%%%%%%%%%%%%%%%%%%%%%%%%%%%%%%%%%%%%%%%%%%%%%%%%%%%%%%%%%%%%%%%
 
\vfil \end{document}